\PassOptionsToPackage{dvipsnames, table}{xcolor}
\documentclass[sigconf]{acmart}
\usepackage{enumitem}
\usepackage{caption}
\usepackage{subcaption}
\usepackage{pifont} 
\usepackage{xspace}
\usepackage{subfiles}
\usepackage{xpatch}
\usepackage{xr} 
\usepackage{hyperref}
\newcommand\sysname{PolarDB-IMCI\xspace}

\usepackage{libertine}
\usepackage{multirow, booktabs}
\usepackage{siunitx}
\usepackage{tabu}%
\usepackage{makecell}
\usepackage{longtable}
\usepackage{tikz}

\definecolor{darkblue}{rgb}{0.0, 0.0, 0.55}
\newcommand{\chref}[1]{\hyperref[#1]{\color{darkblue}\S\ref{#1}}}
\newcommand{\fig}[1]{\hyperref[#1]{\color{darkblue} Figure~\ref{#1}}}
\newcommand{\tab}[1]{\hyperref[#1]{\color{darkblue} Table~\ref{#1}}}

\AtBeginDocument{%
  }

\setcopyright{rightsretained}
\copyrightyear{2023}
\acmYear{2023}
\acmDOI{10.1145/3589785}

\acmConference[SIGMOD'23]{International Conference on Management of Data}{June 18--23,
  2023}{Seattle, WA, USA}
\acmPrice{0}
\makeatletter
\gdef\@copyrightpermission{
  \begin{minipage}{0.2\columnwidth}
  \href{https://creativecommons.org/licenses/by/4.0/}{\includegraphics[width=0.90\textwidth]{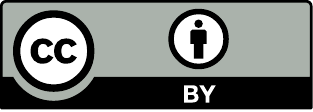}}
  \end{minipage}\hfill
  \begin{minipage}{0.8\columnwidth}
  \href{https://creativecommons.org/licenses/by/4.0/}{This work is licensed under a Creative Commons Attribution International 4.0 License.}
  \end{minipage}
  \vspace{5pt}
}
\makeatother

\begin{document}

%\subfile{revision_letter.tex}

\title{\sysname: A Cloud-Native HTAP Database System at Alibaba}

\author{Jianying Wang$^{\dagger}$, Tongliang Li$^{\dagger}$, Haoze Song$^{\dagger}$, Xinjun Yang$^{\dagger}$, Wenchao Zhou$^{\dagger}$, Feifei Li$^{\dagger}$\\
Baoyue Yan$^{\dagger}$, Qianqian Wu$^{\dagger}$, Yukun Liang$^{\dagger}$, Chengjun Ying$^{\dagger}$$^\S$, Yujie Wang$^{\dagger}$, Baokai Chen$^{\dagger}$ \\
Chang Cai$^{\dagger}$, Yubin Ruan$^{\dagger}$, Xiaoyi Weng$^{\dagger}$, Shibin Chen$^{\dagger}$, Liang Yin$^{\dagger}$, Chengzhong Yang$^{\dagger}$, Xin Cai$^{\dagger}$, \\
Hongyan Xing$^{\dagger}$, Nanlong Yu$^{\dagger}$, Xiaofei Chen$^{\dagger}$, Dapeng Huang$^{\dagger}$, Jianling Sun$^{\dagger}$$^\S$}

\affiliation{%\vspace{0.1cm}
\institution{~\footnotemark[2]Alibaba Group \;\;\;\;\; ~\footnotemark[4]Zhejiang University}
}

\affiliation[ ]{\textit {\{beilou.wjy, litongliang.ltl, songhaoze.shz, xinjun.y, zwc231487, lifeifei, baoyue.yby, daisy.wqq, liangyukun.lyk
                           \\zhencheng.wyj, baokai.cbk, caichang.cc, yubin.ryb, echo.wxy, wuha.csb, allen.yinl, chengzhong.ycz
                           \\frank.cx, diane.xhy, nanlong.ynl, chenxiaofei.cxf, wuzang.hdp\}@alibaba-inc.com}}                        
\affiliation[ ]{\textit {\{yingcj,sunjl\}@zju.edu.cn}}   

\renewcommand{\shortauthors}{Jianying Wang et al.}

\begin{abstract}
    Cloud-native databases have become the de-facto choice for
    mission-critical applications on the cloud due to the need for
    high availability, resource elasticity, and cost
    efficiency. Meanwhile, driven by the increasing connectivity
    between data generation and analysis, users prefer a single
    database to efficiently process both OLTP and OLAP workloads,
    which enhances data freshness and reduces the complexity of data
    synchronization and the overall business cost.

    In this paper, we summarize five crucial design goals for a
    cloud-native HTAP database based on our experience and customers'
    feedback, i.e., transparency, competitive OLAP performance,
    minimal perturbation on OLTP workloads, high data freshness, and
    excellent resource elasticity.
    As our solution to realize these goals, we present \sysname{}, a
    cloud-native HTAP database system designed and deployed at Alibaba
    Cloud.  Our evaluation results show that \sysname{} is able to
    handle HTAP efficiently on both experimental and production
    workloads; notably, it speeds up analytical queries up to
    $\times149$ on TPC-H (100$GB$). \sysname{} introduces low visibility delay and little performance perturbation on OLTP workloads
    (\textless~5\%), and resource elasticity can be achieved by
    scaling out in tens of seconds.

\end{abstract}

  


\maketitle

\section{Introduction}\label{sec:inro}

In recent years, cloud-native
databases~\cite{Aurora,Cao2018,dageville2016snowflake,gupta2015amazon}
have become an inexorable trend in the database industry.  Different
from on-premise databases, a cloud-native database decouples its
architecture into two layers: a computation layer and a storage layer,
allowing resources to scale independently.  Nodes equipped with disks
(in the storage layer) form a shared storage pool that serves as a
unified data interface for nodes in the computation layer.  This
disaggregation architecture enables database systems to offer extreme
elasticity, flexible on-demand charging models, and low operating
costs for customers.  As a result, the market of cloud-native
databases has quickly taken off~\cite{CloudDB}.
% Cloud-native databases are becoming a critical component for data
% management services~\cite{CloudDB}.  For instance,
% PolarDB~\cite{Cao2018}, a cloud-native row-oriented relational
% database, has been widely deployed on Alibaba Cloud, providing
% superb OLTP services for enterprise customers in various
% industries~\cite{li2019cloud}.

\begin{figure}  
  \centering
  \includegraphics[width=\columnwidth]{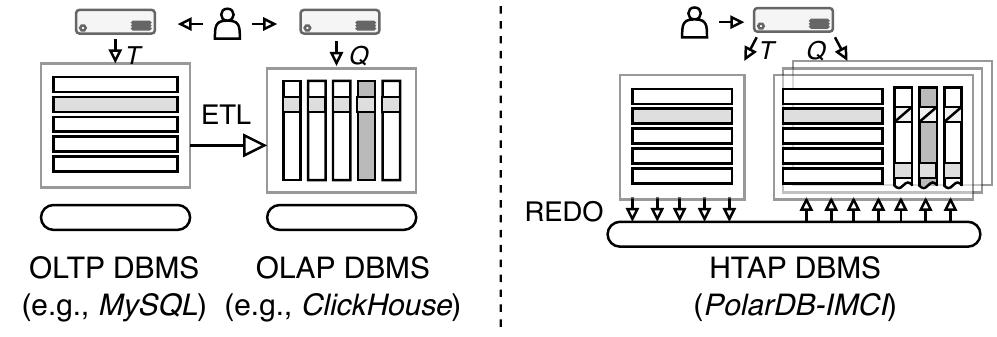}
   \vspace{-2.0em}
  \caption{\small Comparison of ETL and \sysname{}.}\label{fig:intro}
   \vspace*{-1.0em}
\end{figure}

Meanwhile, we have witnessed another trend that the line between
classic OLTP and OLAP databases started to blur: there is a growing
need for a database to provide sufficient support for {\em both}
transactional processing and analytical processing, especially in the
fields of business intelligence~\cite{BI}, social
media~\cite{Twitter,fackbook}, fraud detection~\cite{frauddete}, and
marketing~\cite{marketing,KunPeng}. To provide such capability, traditional
solutions often deploy data and application logic into two databases,
one specialized in OLTP and the other in OLAP (e.g.,
MySQL~\cite{mysql8} for OLTP, and ClickHouse~\cite{ck} for OLAP), and
rely on data synchronization techniques (such as
Extract-Transform-Load~\cite{vassiliadis2009survey} (ETL) workflow)
for ensuring consistencies between them, as shown in \fig{fig:intro}.
According to our statistics, nearly 30\% of the customers of PolarDB,
an OLTP database, synchronize data to an independent data
warehouse system for data analytics needs.

Such solutions are costly, as it negatively impacts the OLTP
performance, and introduces a time-consuming data synchronization
process, which further leads to delays or even inconsistencies between
the data maintained at the TP/AP databases. In practice, these issues
lead to sub-optimal user experience and a large number of user
inquiries. To address these issues, it calls for a cloud-native Hybrid
Transactional and Analytical Processing (HTAP) database. In this
paper, we present \sysname{}, a cloud-native HTAP database deployed at
Alibaba Cloud.
% \sysname{} leverages PolarDB~\cite{li2019cloud} for transaction processing, and proposes several new techniques.
We summarize the crucial design goals of \sysname{} below, 
which are also applicable to the design of a general cloud-native HTAP database.

\begin{itemize}[leftmargin=*]

  \item \textbf{G\#1: Transparent Query Execution.} To serve mixed workloads in
    a single database, database users should not be required to
    understand the working logic of the database, nor should they
    identify query types manually.  That is, users should not
    perceive two {\em isolated} systems (e.g., engines, indexes,
    interfaces, etc.) for OLAP and OLTP queries
    respectively. Our system should provide a unified SQL interface
    for both OLAP and OLTP workloads.
% Note that, it can be extremely challenging to explicitly distinguish OLTP and OLAP queries for complex data-intensive applications. 

\item \textbf{G\#2: Advanced OLAP Performance.}  As a major goal of
  any HTAP database, the OLAP performance (e.g., execution latency) of
  \sysname{} should be comparable to typical databases specialized in
  processing OLAP queries (typically through the introduction of
  columnar data storage).

\item \textbf{G\#3: Minimal Perturbation on OLTP Workloads.}  While the
  performance of OLAP queries is significantly improved, it should
  have a minimal negative impact on the performance of OLTP queries. In
  fact, as we have practically validated in real application
  scenarios, OLTP queries are usually more mission-critical and are
  more sensitive to performance degradation. This requires effective
  resource isolation for OLTP and OLAP queries.
% The dual-format design is not a silver bullet. 
% Columnar data does not replace row-based storage, but supplements it.
% Different from a general-purpose performance isolation~\cite{li2022htap, milkai2022good}, 
% we treat OLTP is much more mission-critical (validated from our customers). 
  % Hence, data movement between two data formats should have the lowest performance impacts on transaction processing.
  % We design PolarDB with IMCI to deliver excellent OLAP-OLTP performance isolation.
  \item \textbf{G\#4: High Data Freshness.}  High data freshness is an
    important property of HTAP databases, which is a distinguishing
    advantage compared to the traditional Extract-Transform-Load (ETL)
    method. In this paper, we follow earlier similar
    work~\cite{tidb2020,chen2022bytehtap} using the visibility delay
    as a freshness score for a query.  By definition, the visibility
    delay is the time interval during which updates to the database
    can be visible to OLAP queries.
  % Based on the feedbacks from our customers, we argue that the data freshness needs to be 
  % less than 200 milliseconds to satisfy the real-time requirements in diverse applications.
  \item \textbf{G\#5: Excellent Resource Elasticity.}  In HTAP
    scenarios, the consumption of CPU/IO resources fluctuates
    significantly, from hundreds to thousands of times. As a key feature
    of cloud-native databases, our system should ensure high resource
    elasticity (e.g., scale-out in minutes or
    even seconds) to adaptively serve the changing data volume and
    analytical workloads with stable performance and high resource
    utilization.
  % \item \textbf{G\#5: Full Transparency.} 
  % The design of IMCI should be transparent to the applications, so that they do not need to be aware of the existence of IMCI. 
  % For example, the applications should not be aware of the existence of IMCI, and they should not need to 
  % explicitly specify the usage of IMCI in their queries.
  % The applications should not be aware of the existence of IMCI, and they should not need to 
  % explicitly specify the usage of IMCI in their queries.
  % \item \textbf{G\#6: Strong Compatibility.}
  % Since PolarDB is a fork of the MySQL~\cite{todo}, 
  % it is completely compatible with the MySQL ecosystem. To integrate IMCI with PolarDB,
  % queries that operate on columnar data must also be MySQL-compatible.
  % Therefore, connecting an external column-oriented engine to PolarDB for analysis queries 
  % may not be the best option.
  % \item \textbf{G\#6: High Availability of Column-store.} High availability is severely important in OLTP databases. 
  % In HTAP, developers rely on this property to safely mix the usage of transactions and analytical queries 
  % or embed both OLTP and OLAP requests in a single transaction. 
\end{itemize}

\sysname{} meets all desired goals (i.e., G\#1-5) with the following innovations. 
First, to meet G\#1 and G\#2, we implemented \textit{in-memory column index (IMCI, \chref{sec:column})} as complementary storage. 
% Typically, OLAP queries access only a subset of attributes with massive rows, thus columnar access reduces the overall I/O cost. 
\sysname{} absorbs diverse advanced optimizations from the OLAP community
% , including batch iteration, parallel operators, SIMD, etc.  
% Based on these optimizations, \sysname{} 
and derives \textit{a new SQL engine (\chref{sec:olap:exe})} to match the execution mode on columns. 
Further, \sysname{} proposes \textit{a new query routing mechanism (\chref{sec:olap:routing})} that dispatches queries transparently.

Second, to meet G\#3, \sysname{} resides column indexes on \textit{separated read-only (RO) nodes (\chref{sec:overview:polar})} with a shared storage architecture to provide effective resource isolation between OLTP and OLAP requests. 
Updates are propagated to RO nodes by \textit{reusing REDO logs (\chref{subsec:logParse})} (i.e., the differential logging for the row store) instead of shipping additional logical logs (i.e., MySQL Binlogs). 
% Such design crucially avoids the need of recording logical operations which could cause non-negligible impact on the OLTP performance. 
% \sysname{} then generates logical operations from REDO logs on the OLAP side.
%  to update heterogeneous IMCI.

Third, to meet G\#4, we enhance our update propagation framework with \textit{commit-ahead log shipping (CALS, \chref{subsec:CALS})} and \textit{2-Phase conflict-free log replay (2P-COFFER, \chref{subsec:coffer})}.
CALS ships transaction logs before committing. 
% REDO logs are buffered at OLAP side, changed to logical ones, and are ready for replaying once the transactions are committed.
2P-COFFER efficiently parses and applies REDO logs to RO nodes.
% , thus multiple replayer can work in parallel without additional overhead on cross-threads coordination. 
Furthermore, we implemented the column index as \textit{append-only storage (\chref{sec:column})}: records are organized in insert order rather than primary key order. 
Thus, updates to column indexes are performed out-place and quickly.
% , favor for data freshness.

Finally, to meet G\#5, the checkpoint mechanism of the columnar index is seamlessly built into PolarDB's original storage engine. Therefore fast scale-out capability can be achieved by quickly pulling up a RO node using the checkpoint on shared storage (\chref{sec:fast_startup}).
%we entend the existing checkpoint mechanism to fully integated with \sysname{}'s column index (\chref{sec:fast_startup}).

% (3). Data changes to IMCI are performed by replaying redo logs placed in shared storage 
% without data migration (to meet \textbf{G\#4}). 
% New records are appended  into a delta store, while expired records are logically deleted by marks. 
% Additionally, a new crash recovery mechanism is introduced in IMCI (to meet \textbf{G\#6}).
% Hence, maintaining column indexes is never a bottleneck for committing transactions.

% In this paper, we share our experience on developing \sysname{}. 
We started the design and development of cloud-native PolarDB in 2017, and seek for an HTAP solution (i.e., \sysname{}) in 2019. 
By now, \sysname{} is severing a large number of internal and external customers (\tab{tab:speedup}).
The key contributions of this work are listed as follows:
\begin{itemize}
  \item We propose \sysname{}, an HTAP solution for cloud-native relational database systems.
  To the best of our knowledge, \sysname{} is the first cloud-native HTAP database 
  to satisfy all of the aforementioned design goals.
  \item We design an architecture that provides dual-format storage on read-only nodes under the storage-computation separation architecture, 
  which enables efficient execution of analytical queries and minimizes the impact on OLTP load. Additionally, 
  \sysname{} is the first practical template to demonstrate that it is possible and applicable to implement replication from row-store to dual-format storage with physical redo logs while reducing replication latency to millisecond levels.
  % HZ: 
  % \item \textcolor{blue}{We present several key techniques to collectively
  % make a fresh, high-performance column index with advanced vectorized parallel execution engine for analytical queries. 
  % Additionally, \sysname{} is the first practical template to to demonstrate that it is possible and applicable to synchronize heterogeneous store with physical logs in HTAP scenarios.}
  %  We share the lessons learned in implementing \sysname{}, and take the first step to demonstrate it is possible to synchronize heterogeneous store with physical REDO logs.
   \item We evaluate \sysname{} with diverse experiments (in both experimental and production environments).
  The experimental results show that \sysname{} outperforms row-based PolarDB up to $\times149$ on a standard analytical workload TPC-H (100$GB$), 
  and its performance is comparable to the advanced OLAP databases (e.g., ClickHouse). 
  Performance degradation on OLTP is tiny (less than 5\%), even when OLAP workloads increase continuously.
  The visibility delay of \sysname{} at is \textless$5ms$ on typical workloads, and \textless$30ms$ under heavy workloads. \sysname{} can scale out in tens of seconds.

  % and its performance on TPC-H is comparable to the advanced OLAP databases (e.g., Clickhouse).
  %  We further indicate that the OLTP performance degradation caused by IMCI is minimal (less than 10\%).
\end{itemize}

The remainder of the paper is organized as follows. 
\chref{sec:backg} introduces the background of HTAP and cloud-native databases. 
\chref{sec:overview} presents the architecture. 
\sysname{}'s components and update propagation framework are introduced in \chref{sec:column} and \chref{sec:upd} respectively.
\chref{sec:olap} discusses query dispatch, optimization, and execution. 
\chref{sec:fast_startup} introduces the checkpoint mechanism. 
\chref{sec:eval} details the experiments and evaluation. 
\chref{sec:conclusion} concludes the paper.

\section{Background and Related Work}
\label{sec:backg}
\subsection{Hybrid Transactional/Analytical Processing}

For long decades, OLTP and OLAP databases are dedicatedly designed for their respective workloads. For instance, OLTP engines (e.g., \textit{MySQL}~\cite{mysql8}) prefer row-based data formats, row-at-a-time operators, and early materialization strategy, favoring data modification and point queries. On the contrary, OLAP engines (e.g., ClickHouse~\cite{ck}) use column-based data formats, batch-at-a-time operators, and late materialization strategy, favoring scan-intensive analytical queries. As a result, modern database administrators often need to deploy both OLTP and OLAP databases, and conduct data shipping between two types of databases (e.g., ETL~\cite{vassiliadis2009survey}).

The emergence of HTAP databases eliminates the burden of maintaining multiple databases and simplifies data shipping.
% Over the past decade, many systems~\cite{} in the HTAP landscape have been proposed.  
We classify existing HTAP solutions into two categories (i.e., single-instance and multi-instance),
and discuss each category below.

\vspace{1.5mm}
\noindent\textbf{HTAP with Single Instance.} 
SAP HANA~\cite{hana2012} supports hybrid workloads by introducing a
three-tier merge tree, a layered in-memory store that supports both
row and column formats.
Oracle Dual~\cite{oracle2015} allows relational tables to be built 
as In-Memory Column Units (IMCU) to provide fast column scans.
New updates to IMCUs are temporarily logged by metadata, 
and IMCUs can be repopulated from the memory buffer when more updates are accumulated.

Unlike Oracle Dual, SQL Server CSI~\cite{larson2011sql, larson2015real} supports column stores with column store index (CSI) and periodically merges new updates into CSI, thus eliminating rebuilding.
% Merges result in row movements, so an additional mapping mechanism is required to look up a matching row.
% When a column store is used as a secondary index, the row-store table has a hidden row ID (RID) attribute 
% that keeps track of  a row's location in the column store.
% When a column store is used as the base storage, SQL server uses an auxiliary mapping index 
% to avoid row movements affect multiple secondary indexes.
\sysname{} follows a similar principle, but
pioneers this design to the cloud-native architecture by addressing a
number of key challenges as detailed in later sections.
%SingleStore~\cite{memsql2022} provides an in-memory row store, and merge data into an on-disk column store continuously. 

%Different from these works, \sysname{} implements its in-memory coloumn store as a secondary index (i.e., IMCI).  
%Since new updates to IMCI is append-only and out-place (\chref{sec:column:layout}), IMCI can be upated in real-time,
%eliminating the rebuilding or merge process.

% , initially designed as an in-memory columnar database for purely OLAP applications, 
% has continued to advance toward mixed workloads. 
% To reduce the impact of transactional queries on analytical queries, 
% SAP HANA adopts a three-tier merge tree architecture to absorb data updates.

\vspace{1.5mm}
\noindent\textbf{HTAP with Multiple Instance.}
Another type of HTAP database utilizes replication techniques to maintain multiple instances.
Thus, transactional and analytical queries can be routed to different instances to achieve efficient performance isolation. 
Further, each instance can tailor its architecture to fit workloads.
%To minimize the performance perturbation on OLTP, we deploy \sysname{} with multiple instances 
%by default (see \chref{sec:overview:polar}).

% In a typical architecture, replicas serve OLAP workloads 
% while the primary instance handles OLTP workloads.
A more recent work of SAP HANA proposes Asynchronous Table Replication (ATR)~\cite{lee2017parallel} for 
data synchronization between the primary instance and replicas.
Replication logs are supplied asynchronously to replicas 
and are replayed in parallel in session granularity. Unlike ATR, 
Google F1 Lightning~\cite{f12020} uses Change Data Capture (CDC), 
a more loosely coupled mechanism shuffling data via BigTable.
% Despite having less freshness than ATR, CDC allows users to achieve good HTAP performance
% without moving out data from in-use OTLP systems.
TiDB~\cite{tidb2020} uses Raft~\cite{ongaro2014search} to connect row-store engines (TiKV) and columnar engines (TiFlash).
TiFlash behaves as a Raft learner receiving logs asynchronously from the leader 
and does not participate in the leader election. IBM DB2
Analytics Accelerator (IDAA)~\cite{butterstein2020replication} maintains a copy of row-based table data by integrated synchronization to
support incremental updates. 
A new version of Oracle Dual~\cite{pendse2020oracle} supports offloading read-only workloads to homogeneous instances (standby) and synchronizes data by REDO logs.  
ByteHTAP~\cite{chen2022bytehtap} uses disaggregated storage and synchronizes heterogeneous engines (ByteNDB for OLTP and Apache Flink~\cite{flink} for OLAP) by Binlog.
Wildfire~\cite{barber2016wildfire} is a Spark-compatible database and also leverages disaggregated storage for data synchronization. 
Different from these works, \sysname{} directly reuses REDO logs for heterogeneous data replication. To the best of our knowledge, \sysname{} is the first industrial database using physical logs to efficiently synchronize heterogeneous storage.

% in a Raft group for transactional queries.
% TiDB enhances HTAP performance by adding learners, a new role in Raft.
% A learner is 
% a columnar engine (TiFlash)  and receives logs sent asynchronously from the leader.
% The learner mechanism brings benefits as well as drawbacks. 
% On the one hand, TiFlash does not participate in Raft protocol (e.g., log commits or leader election) to reduce the overhead of leaders. 
% On the other hand, asynchronous log replication results in a visible latency of more than 1 second.

%Additionally, several databases leverage shared storage for data synchronization.
%Wildfire~\cite{barber2016wildfire} is a Spark-compatible database. 
%It uses SparkSQL as the engine for analytical processing.
%Data updates are first committed to the local SSD and then moved asynchronously to its shared storage. 
%\sysname{} also adopts a shared storage (i.e., PolarFS~\cite{Cao2018}), but supports real-time synchronization.

\subsection{Cloud-Native Database}

% The adaption of cloud-native database is proved to be successful in multiple scenarios.
The key technique of cloud-native databases is decoupling computation and storage. 
A typical cloud-native database often adopts cloud storage underneath
its storage engine, leveraging another layer for virtualization and
providing an elastic storage service~\cite{chen2022cloudjump}. 
Cloud-native architecture benefits customers with high resource elasticity and an on-demand charging model 
and benefits service providers by reducing maintenance and development costs.
% It helps reduce the maintenance cost and expedite development cycles for the database kernels.

\vspace{1mm}
\noindent\textbf{Cloud-native OLTP/OLAP.} 
Aurora~\cite{Aurora, verbitski2018amazon} is a cloud-native OLTP database deployed on a custom-designed cloud storage layer. 
Taurus~\cite{depoutovitch2020taurus} uses asymmetric replication based on separate
persistence mechanisms for database logs and pages. 

Besides OLTP systems, OLAP databases also benefit from
storage-disaggregation.  Several conventional data warehousing systems
have adapted to the cloud (e.g., Vertica~\cite{lamb2012vertica},
Eon~\cite{vandiver2018eon}), and several OLAP databases are natively
developed for the cloud (e.g., Snowfake~\cite{dageville2016snowflake},
Redshift~\cite{gupta2015amazon}, AanlyticDB \cite{DBLP:journals/pvldb/ZhanSWPLWCLPZC19}).

\vspace{1mm}
\noindent\textbf{Cloud-native HTAP.} SingleStore~\cite{memsql2022} takes the first step to make the HTAP database cloud-native. 
It disaggregates computation and storage, and supports committing transactions on the local disk of computation nodes and pushing
data asynchronously to its blob storage. Different from SingleStore, \sysname{} offloads all persisted data into the shared storage layer, 
thus all states of the computation nodes can be rebuilt from shared storage directly, favoring recovery and elasticity.

\section{Overview}\label{sec:overview}

In this section, we first outline the architecture of
\sysname{}, then summarize the design rationales driven by the
aforementioned design goals, along with a brief description of the user interface.

\subsection{Architecture of \sysname{}}\label{sec:overview:polar}

\begin{figure}  
  \centering
  \includegraphics[width=1\columnwidth]{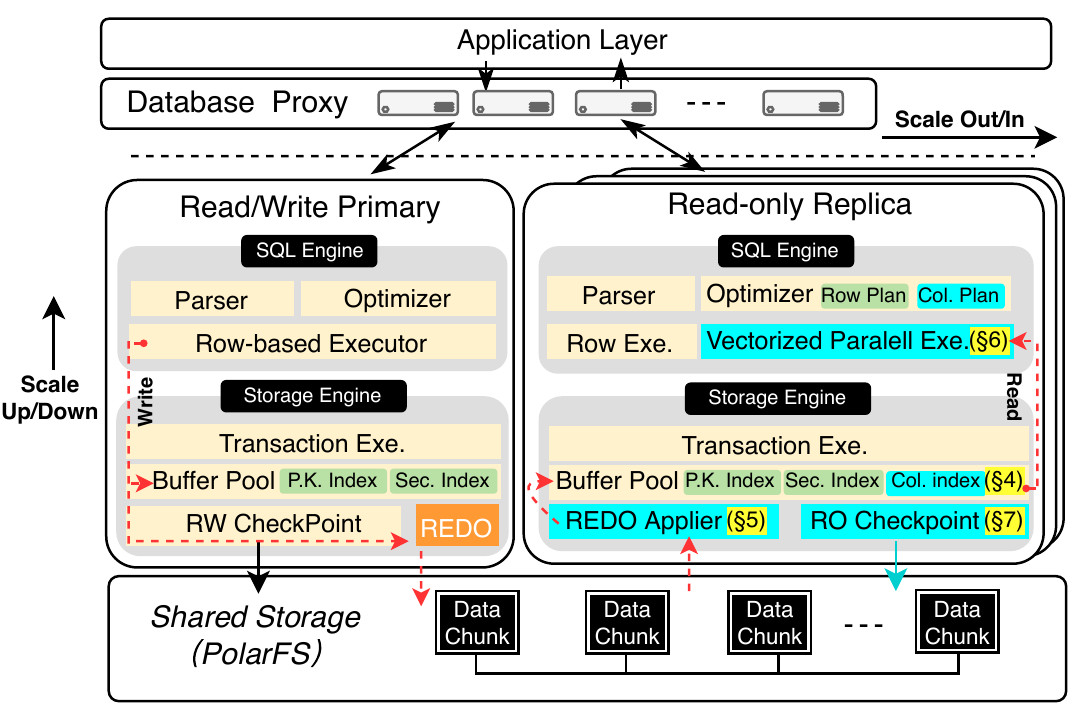}
  \caption{\small Cloud-native architecture of \sysname{}. } \label{fig:architecture}
  \vspace*{-10pt}
\end{figure}

\fig{fig:architecture} shows the architecture of \sysname{}, which
follows the crucial design principle of separating computation and
storage architecture. The storage layer is a user-space
distributed file system called PolarFS~\cite{Cao2018} with high
availability and reliability.  The computation layer contains multiple
computation nodes, including a primary node for read/write requests
(RW node), several nodes for read-only requests (RO nodes), and
several stateless proxy nodes for load balancing.  Given this,
\sysname{} can provide high resource
elasticity (\chref{sec:fast_startup}).  Furthermore, all nodes in both
storage and computation layers are connected by a high-speed RDMA
network to achieve low latency of data access.

To speed up analytical queries, \sysname{} supports building in-memory
column indexes (\chref{sec:column}) on the row store of RO nodes.
Column indexes store data in insertion order and perform out-place
writes for efficient updates.  The insertion order means a row in
column indexes that can be quickly located by its Row-ID (RID) rather
than its primary key (PK).  To support PK-based point lookups,
\sysname{} implements a RID locator (i.e., a two-layer LSM tree) for
PK-RID mapping.

\sysname{} uses an asynchronous replication framework
(\chref{sec:upd}) for synchronization between RO and RW.  That is,
updates to RO nodes are not included in the transaction commit path of
the RW to avoid the impact on the RW node.  To enhance data freshness
on RO nodes, \sysname{} uses two optimizations on the log applying,
the commit-ahead log shipping, and the conflict-free parallel log replay
algorithm.
RO nodes are synchronized by REDO logs of the row store,
which causes very low perturbation on OLTP than other strawmen
approaches (e.g., using Binlog). Note that it's nontrivial to apply
physical logs into column indexes as the data format of the row store
and column index is heterogeneous.

Inside each RO node, \sysname{} uses two mutually symbiotic execution
engines (\chref{sec:olap}): PolarDB's regular row-based execution
engine to serve OLTP queries, and a new column-based batch mode
execution engine for the efficient running of analytical queries.  The
batch mode execution engine draws on the techniques used by columnar
databases to handle analytical queries, including a pipeline execution
model, parallel operators, and a vectorized expression evaluation
framework.  The regular row-based execution engine with augmented
optimizations can undertake the column engine's incompatible queries
or point queries. 
\sysname{}'s optimizer automatically generates and coordinates plans for both execution engines, which is transparent to the consumer.

\subsection{Design Rationales}
We highlight the design rationales of \sysname{} below, which may also apply to other cloud-native HTAP databases.

\noindent\textbf{Storage-Computation Separation.} As a key design principle of
cloud-native databases, the storage-computation separation architecture enables adaptive compute resource provisioning to shifting workloads without data movement, which has become a mainstream architecture alternative. \sysname takes the decision to naturally match our design goal \textbf{G\#5} (high resource elasticity).

\vspace{1mm}
\noindent\textbf{Single RW Nodes with Multiple RO Nodes.} 
As a practical design decision, single-writer architecture has been confirmed to have advanced write performance~\cite{Aurora} and significantly reduce the system complexity. We have observed that a single RW node is enough to serve 95\% customers in our business. With the design choice, all RO nodes have a consistent data view synchronized with the RW node. Large OLAP queries are routed to RO nodes to enable efficient resource isolation and the RO nodes can be quickly scaled out to serve surging OLAP queries, which follows the design goal \textbf{G\#3} (minimal perturbation on OLTP) and \textbf{G\#5} (resource elasticity).

\vspace{1mm}
\noindent\textbf{Hybrid Execution and Storage Engines inside RO Nodes.}  
From the lessons in the OLAP community, columnar data layout and vectorized batch execution are significant optimizations for OLAP queries. 
However, it is not a wise decision for us to use an existing column-oriented system (e.g., ClickHouse) to serve directly as RO nodes.
There are two reasons for this.
First, it is time-consuming to achieve full compatibility between the RW node and RO nodes.
In a cloud service environment, even little incompatibility can be drastically amplified and overwhelm developers given the huge customer volume.
Second, pure column-oriented RO nodes are still inefficient for
point-lookup queries, which are classified as OLTP
workloads.  As a result, we started to design a new column-based execution engine extending the original execution engine of PolarDB,
to satisfy the goal \textbf{G\#1} (transparency). 
The column-based execution engine is designed to meet \textbf{G\#2} (advanced OLAP performance). 
While the row-based execution engine handles incompatible and point-lookup queries that the former cannot deal with.
RO nodes have both column-based and row-based execution and storage engines.

\vspace{1mm}
\noindent\textbf{Dual-format RO Nodes Synchronized by Physical REDO Logs}. 
  With the architecture over the shared storage, new RO nodes can be quickly started to serve surging read-only queries to meet the design goal \textbf{G\#5}, 
  and can continuously apply REDO logs from the RW node to keep storage fresh (i.e., \textbf{G\#4}). 
  However, synchronizing heterogeneous storage with the original physical logs (i.e., REDO logs) is challenging as the logs are tightly coupled 
  with the underlying data structures (e.g., pages).
  Therefore, a strawman approach is letting the RW node record additional logical logs (e.g., Binlog) for column-store. The drawback is significant: 
  it triggers additional \textit{fsyncs} when committing a transaction, thus causing non-negligible performance perturbation on OLTP.
  Given this, we dedicatedly designed a new synchronization method by reusing REDO and making up logical operations from physical logs on RO nodes.
  It is feasible since \sysname{} maintains both row-based buffer pool and column indexes on RO nodes. 
  Logical operations can be regained by the applying process on a row-based buffer pool.
  Our evaluation shows that the overhead of reusing REDO logs is significantly lower than using Binlog.

  \begin{figure}  
    \centering
    \setlength{\belowcaptionskip}{-15pt} 
    \setlength{\abovecaptionskip}{5pt} 
    \includegraphics[width=\columnwidth]{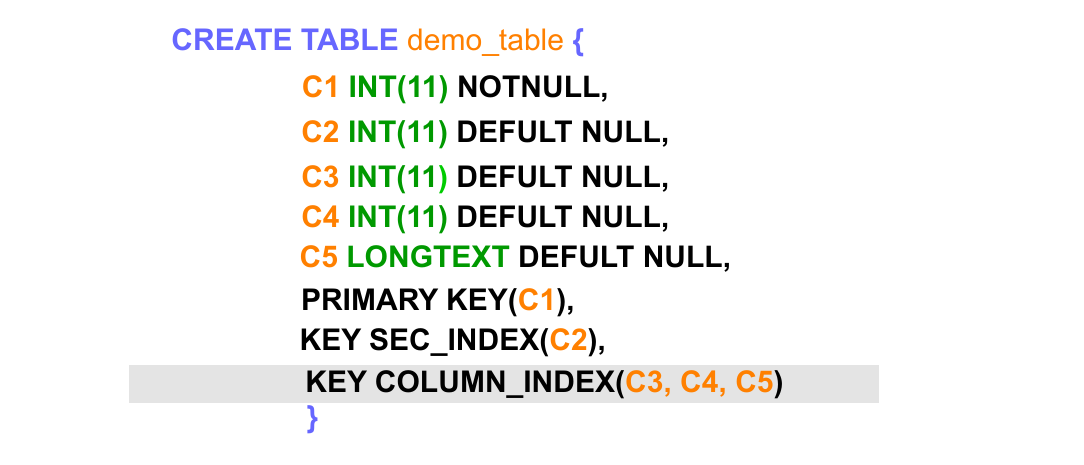}
    \caption{\small A DDL creates a demo table with a primary key index on
      C1, a secondary index on C2, and column indexes on column
      C3,C4,C5.} \label{fig:sql}
  \end{figure}

\subsection{User Interface}
Column store in \sysname{} is exposed as a new index type: column
index. Applications can create a column index for a table on
demand. As \sysname{} is fully compatible with \textit{MySQL},
applications can use the SQL statement with \textit{MySQL} syntax to
create a column index. An example is shown in \fig{fig:sql}. It
creates a table with five columns, the primary key index is created on
column C1, a secondary index is created on column C2, and column
indexes are created on columns C3, C4, and C5.

In addition, to specify the columns included in column indexes when
creating the table, applications may also use the ALTER statement to
add a column index later. When applications execute Data Definition
Language (DDL) on a table with a large number of rows to add a column
index. The RO node will issue a consistent read on \sysname{}'s
row store, scan the checkpoint, and convert it to a column index in
parallel. Note that adding column indexes in \sysname{} is an online
operation: the queries and DML operations on the table can process
together while a DDL operation is in progress. The changes made by
concurrent DML operations will be recorded in a buffer and applied to
the new column index at the end of the process.

\section{Column Index Storage} \label{sec:column}
% PolarDB supports two types of engines, namely the B-Tree storage engine and the LSM-tree storage engine X-Engine \cite{xengine2019}. Both of them are row-oriented. 
%The B-tree storage engine is used to support high-performance transaction processing scenarios. The LSM-tree storage engine has a high compression ratio and is used for data archiving scenarios.
%In the B-tree storage engine, a table always has one primary index and multiple secondary indexes and the primary index can only be B-tree. 
%Column index implements the same access interface as a common index. Optimizers and operators can use column index like ordinary B-tree indexe s such as scan/point lookup. 
%After outlining the overall architecture and design decisions of \sysname{}. 
%In this section we explain the storage architecture in detail. 
%\sysname{} oargnizes column data in insertion order which improves write performance. 
%It also uses a two-layered LSM-tree to map the primary key to the 
%physical position of the row in the column index which can support fast delete and update. 
This section dives into the column index store, a crucial part of \sysname{} for handling analytical queries. \sysname{} supports row-based storage engines~\cite{chen2022cloudjump,Huang2019} that are highly tuned for transaction processing on cloud storage. However, row-based data formats are well known for being inefficient to serve analytical queries. Inspired by pioneering industrial databases (e.g., Oracle~\cite{oracle2015}, SQL Server~\cite{larson2015real}), \sysname{} implements a dual data format via in-memory column indexes, to enhance OLAP functionality.

\subsection{Data Organization of Column Index}\label{sec:column:layout}

\begin{figure}  
  \centering
  \includegraphics[width=1\columnwidth]{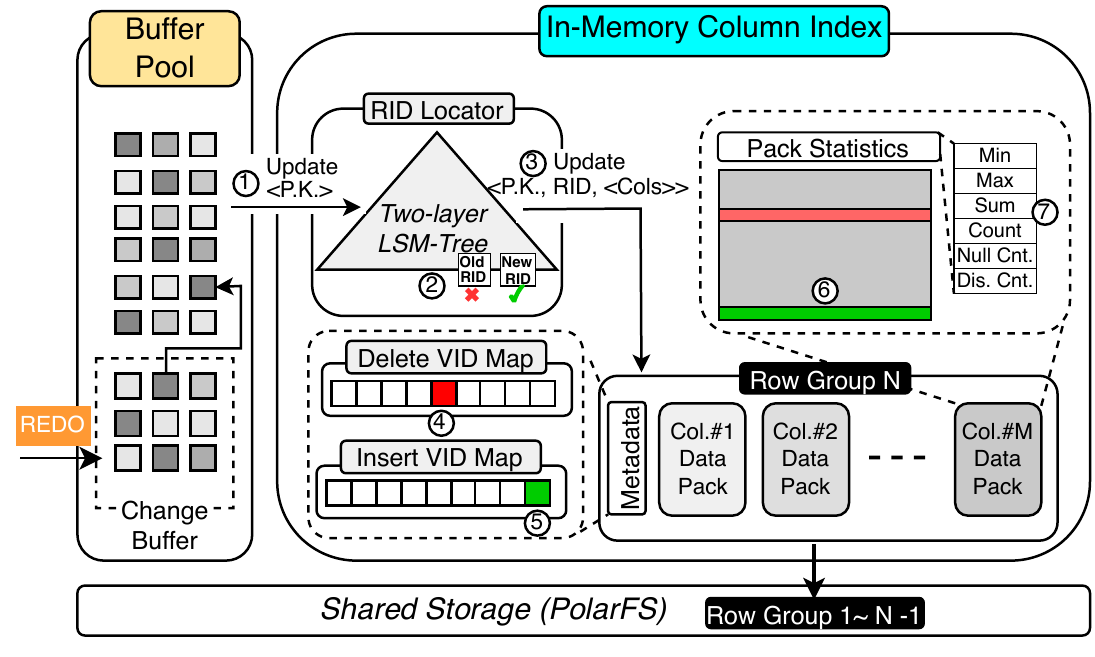}
  \vspace{-2em}
  \caption{\small This diagram shows how data is updated in IMCI storage (i.e., step \ding{172}$\sim$\ding{178}). 
  For simplicity, both delete and insert operations are performed in the last column data pack (i.e., partial Packs). 
  ``RID'' means row id. ``VID'' means version id.} \label{fig:IMCI}
  \vspace{-1em}
\end{figure}

As shown in \fig{fig:IMCI}, column indexes in \sysname{} serve as complementary storage to the existing row store. In \sysname{}, columns of a table can selectively be involved in a column index. \sysname{} divides all rows of a table into multiple row groups with append-only writes to improve the write performance. In a row group, each column of data is organized into a data pack, along with some metadata for statistics. To provide snapshot isolation, each row group contains an insert Version Id (VID) map, and a delete version ID map to control the visibility for concurrent transaction processing. Since the row groups are append-only, deletes require an explicit row id for the given primary key to set the delete version for that row. To realize it, \sysname{} implements a Row-ID locator (i.e., a two-layered LSM tree) to map the primary key to the physical position of the row in the column index.

\noindent\textbf{Data Pack Layout.}
A relational table is first divided into multiple row groups with configurable size (i.e., 64K rows per row groups), and the left rows form a partial row group (e.g., Row Group N in \fig{fig:IMCI}). To realize fast data ingestion, row groups are append-only (\chref{sec:column:dml}). 
That is, the full-sized row groups are immutable, and partial row groups will be fulfilled in an append-only manner. The data belonging to the same column within a row group is organized as a Data Pack in a compressed format to reduce space consumption. Note that \sysname{} does not compress Partial Packs as they are updated continuously. 
% When a partial pack is full, it gets compressed into a pack.
%Insertions are performed by appending a row directly to partial packs without moving other rows, which results in advanced write performance. Deletions simply mark a row as invalid to logically delete it and update was converted to a delete with a insert. As a result, column indexes are arranged in insertion order.

% More specifically, there are two type of data packs: fixed-length and variable-length. 
% In a fixed-length packs, all columns are stored in insert order and can be positioned directly by multiplying the column length by the offset. 
% For variable-length packs, the offset and length of each columnar data are stored at the head of the pack, and the content is stored at the end of the pack. 
% According to the offset and length, the content can be located quickly.

%Different field types have different pack size. The pack size of an 8Byte integer column is 512KB(8*64K). Pack is a larger storage unit which compared with a row-oriented page of 16KB. Large pack size is more conducive to sequential scanning. 

\noindent\textbf{Pack Meta.}
To avoid unnecessary data access during query execution, \sysname{} maintains a Pack meta for each Data Pack. The Pack meta keeps track of minimum and maximum values as well as a sampling histogram for each Pack, which benefits column scan. For instance, when a query statement specifies a WHERE clause predicate, Pack meta for the referenced column can be used to check whether the scan on this Pack can be skipped. 
\iffalse
Furthermore, for those fact tables in data warehouse (where updates are fewer),
\sysname{} supports specifying the whole column index (i.e., all Packs in all row groups) to be sorted according to 
the order of a given column (in alphabetical order) when creating. 
\textcolor{blue}{This maintains the global order of the entire relational table, 
thus the minimum and maximum values of different data Packs (i.e., the ordered data Packs from different row groups) 
do not overlap and can further improve the scan efficiency. }
% Newly inserted rows are still organized in insertion order.
\fi

\noindent\textbf{RID Locator.}\label{sec:column:mapping}
As the data in Packs is stored in its insertion order, \sysname{} relies on a locator to map primary keys to their corresponding physical locations in column indexes. In \sysname{}, each row is assigned an increasing and unique Row-ID (RID) by its insertion order. Then, the RID locator records the mapping of Key-Values pairs (i.e., <Primary Key, RID> ). 
Delete operations rely on the locator to find the physical position of records.
\sysname{} uses a two-layered LSM tree for the RID locator.
%  and does not need to handle large scans or point queries. 
% unordered data structures like hash tables are also an option.  
Compared to other data structures, the LSM tree helps the locator achieve near-optimal memory utilization and easily extends to disks. 
%In our practice, the performance of the RID locator is not the bottleneck (\fig{fig:propagation:AP})
%(\fig{fig:VD:write:tput}).

%\noindent\textbf{Column Index MVCC.}
%The column index uses the well known Multi-Version Concurrency Control (MVCC) to support snapshot isolation. There are an insert VID map and a delete VID map for each row group to maintain the data versions. 
%The insert and delete VID for each row are 64-bit integers to record the committed sequence number of transactions. Since all updates to column indexes are append-only, old versions of VID are always retained. 
%A read query determines whether the data is visible by comparing its version with the row's insert and delete VID. 

\noindent\noindent\textbf{Version Id (VID) Map.}
PolarDB-IMCI uses Multi-Version Concurrency Control (MVCC) to provide consistent data views. 
For column indexes, updating a record is appending a new version of this record to the tail of Partial Packs. 
% Hence, concurrent transactions do not interfere with each other.
Each version has a 64-bit insert VID and a delete VID, recording the timestamps of the appending and deleting of this version, respectively.
The old version of the record is logically deleted by marking it with a timestamp. 
A read transaction determines a version is visible by checking its timestamp is within the range of the insert VID and the delete VID.
% All insert (delete) VIDs in a row group form an insert (delete) VID map.

\subsection{DML Operation on Data Packs}\label{sec:column:dml}
To better understand the process flow on data packs, we now describe how to conduct DML operations on the data structure of column indexes.

\begin{itemize}[leftmargin=*]
\item \textbf{Insert:}
Inserting a row into a column index consists of the following four steps. First, the column index allocates an empty RID from its Partial Packs.
Second, the locator updates the new RID by the primary key for the inserted row (i.e., add a new record into the LSM tree).
Then, the column index writes row data into the empty slot (e.g., data packs within the Row Group N in \fig{fig:IMCI}). 
Finally, the insert VID records the transaction committed sequence number (i.e., timestamp) of the inserted data. Since the insert VID map maintains the insert version of each inserted data, it also follows the append-only write pattern.

\item \textbf{Delete:}
The delete operation retrieves a row's RID via the RID locator by its primary key (PK) and then sets the corresponding delete VID with its transaction committed sequence number. After that, the mapping between the PK and RID is removed from the locator to ensure data consistency.

\item\textbf{Update:}
As shown in \fig{fig:IMCI}, an update on the column index is performed as a delete operation followed by an insert operation. The updated version of a row is appended to Partial Packs, and the old version is logically deleted from its original data Pack (i.e., set the delete VID to max value).
\end{itemize}
%%todo ,updata example?
%In summary, insert operations are performed by appending a row directly into Partial Packs without moving other rows. Delete operations simply mark the delete VID to logically delete them. 
As a result, column indexes are arranged in insertion order with fast data ingestion. Another significant benefit of out-place updates is that it avoids the contention for modification of the same row (\chref{subsec:logrepaly}).
% It is important to note that logs from different rows can be replayed in parallel. 
% Please see the details in \chref{subsec:logrepaly}.

\subsection{Data Pack Compression and Compaction}
\noindent\textbf{Compression.}
A Partial Pack is transformed into a Pack when it reaches its maximum capacity and then compressed into disks to reduce space consumption.
The compression process is carried out with a copy-on-write pattern to avoid access contentions. That is, a new Pack is generated to hold the compressed data, with no changes to the Partial Pack.
\sysname{} updates the metadata after compression to replace the Partial Pack with the new Pack (i.e., atomically updating the pointer to the new Pack). 
For the various data types, column indexes employ different compression algorithms.
Numerical columns adopt the combination of frame-of-reference, delta-encoding, and bit-packing compression, and string columns use dictionary compression.

Additionally, since Packs are immutable, the insert VID map of that Pack is useless when active transactions are greater than all VIDs, i.e., no active transactions refer to the insert VID map. 
In such cases, \sysname{} removes the insert VID maps in row groups to reduce memory footprint.

\noindent\textbf{Compaction.}
Delete operations may set delete VIDs in a Pack, which punches holes for that Pack.
As the number of invalid rows increases over time, the scan performance and the space efficiency degrade. 
\sysname{} periodically detects and re-arranges under-flowing Packs to keep a low waterline for invalid rows of column indexes. 
For example, sparse Packs, with less than half of the valid rows, are picked as under-flowing. 
Then the background threads issue a compaction transaction, which includes numerous update operations, one for each migrated valid row, to re-append all valid rows of picked Packs into Partial Packs. 
Recall that the update operations of column indexes are out-place, so the old rows are still accessible for foreground operations during or even after the compactions, which enables non-blocking updates. 
The picked Packs after compactions will be permanently removed when no active transaction accesses them. 
%Since PolarDB-IMCI maintains a dictionary from Pack-ID to Pack's physical position, a Pack removal is simply performed by deleting the Pack-ID from the dictionary.

%\textcolor{red}{How to locate physical position by row id when a Pack is removed since it is calculated by O(1) direct address mapping}

%Creating column indexes on partial column collections is very helpful.In the HTAP scenario, the schema designed according to the online transaction processing business model often contains some fields of text/blob/json type. But analytical queries often read and process fields of type int/double/decimal. Excluding text/blob fields from column indexes saves memory and storage costs. Users can also easily add or delete columns in the column index according to the changes in Query. 

\section{Update Propagation}\label{sec:upd}

% After outlining the overall architecture and design decisions of \sysname, 
% this section goes into great detail about its update propagation mechanism.
In this section, we describe our efforts for synchronizing heterogeneous data storage.
Minimal perturbation on OLTP is a high-priority goal for \sysname.
To achieve this goal, update propagation in \sysname{} is implemented by REDO logs, 
eliminating the overhead for RW to persist additional logical logs.
On top of REDO logging, PolarDB needs to keep RO nodes as up-to-date as possible for data freshness.
For this purpose, we introduce Commit-Ahead Log Shipping (CALS) to reduce visible delay 
and 2-Phase COnFlict-Free parallEl Replay (2P-COFFER) mechanism to improve replay throughput.

\begin{figure}      
    \centering 
    \includegraphics[width=0.95\columnwidth]{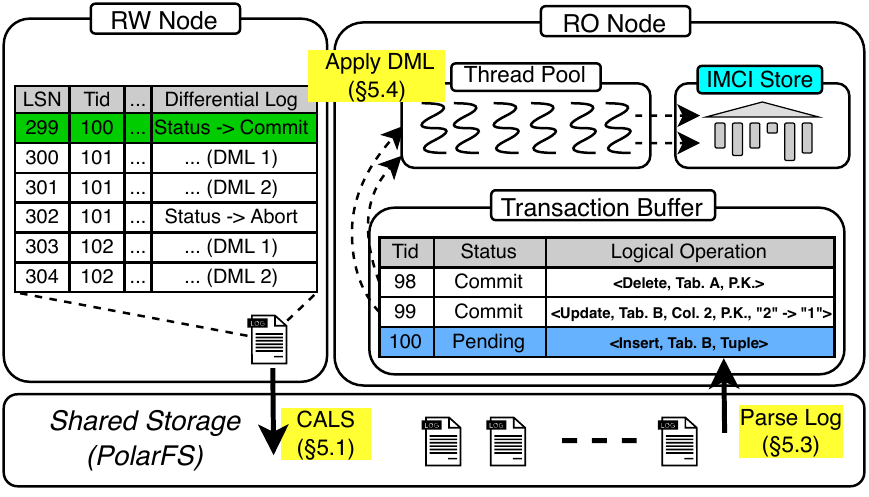}
    \vspace{-1.2em}
    \caption{\small An overview of REDO Log shipping. Logs are shipped from RW node to RO node by shared storage (PolarFS).} \label{fig:shipping}
    \vspace{-1.5em}
\end{figure} 

\begin{figure*}  
    \centering 
    \includegraphics[width=2\columnwidth]{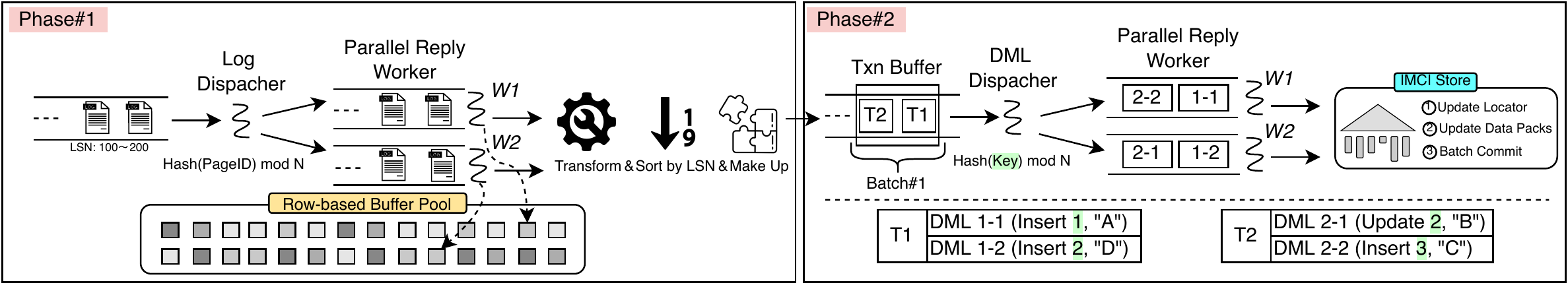}
        \vspace{-1em}
    \caption{\small This diagram shows \sysname{}'s Two-Phase Conflict-Free Parallel Replay (2P-COFFER). 
    In the first phase, REDO logs are parallelly replayed to the row-based buffer pool, transformed to logical DMLs, sorted, and made up to form transactions.
    In the second phase, DMLs (inside each transaction) are processed in batches. DMLs are dispatched based on their primary keys and update column indexes parallelly.} \label{fig:parallel}
    \vspace{-1.2em}
\end{figure*}

\begin{figure}  
    \centering 
    \includegraphics[width=0.95\columnwidth]{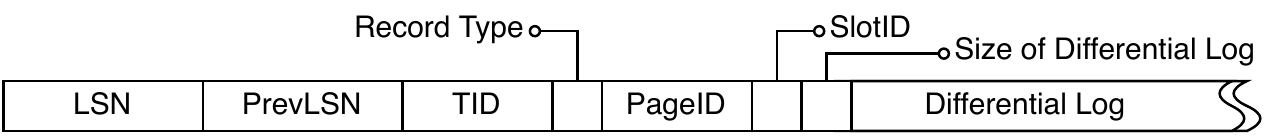}
    \vspace{-0.5em}
    \caption{\small REDO Log Format. } \label{fig:redo}
    \vspace{-1.6em}
\end{figure}

\subsection{Commit-Ahead Log Shipping}\label{subsec:CALS}

% First, we demonstrate how to ship logs to RO nodes under the architecture of \sysname.
% Updates of RO nodes are asynchronous without affecting RW transaction commits.
To minimize performance perturbation, in \sysname{}, updates to RO nodes are fully asynchronous.
% The main challenge for log shipping is how to keep RO nodes as up-to-dated as possible.
Given this, to enhance data freshness, \sysname{} uses the CALS technique, 
which ships transactions before committing.
As illustrated in Figure~\ref{fig:shipping}, a transaction consists of multiple log entries:
the last entry is a commit or an abort log, whereas the ones before it is DML logs.
Each log entry is assigned a log sequence number (LSN). 
For example, the transaction with TID 101 has three log entries with LSN $300$$\sim$$302$. Log entries $300$ and $301$ are DMLs.
Log entry $302$ contains the decisions on the transaction (i.e., abort).

After the RW node writes a log entry to the shared storage (i.e., PolarFS), it notifies RO nodes by broadcasting its up-to-date LSN ($299$ in our example). 
When receiving LSNs, RO reads logs from PolarFS immediately.
Each DML log is then parsed into a DML statement and stored in a transaction buffer 
based on its TID (one buffer unit per transaction).

The whole process does not require waiting for the RW node to commit the transaction. 
For example, the DMLs in the transaction with TID $100$ will ship before the final commit in log entry $299$.
When the RO node reads a commit log entry, 
the earlier DML statements are already parsed and delivered as logical operations in the transaction buffer, allowing \sysname{} to replay the DMLs immediately.
% Note that the RW node may abort transactions.
When reading an abort log entry, RO simply frees the transaction buffer and no data need to be rolled back.

\subsection{Two-Phase Conflict-Free Parallel Replay}\label{subsec:coffer}
As mentioned previously, \sysname{} does not generate additional logical logs for update propagation but reuses REDO logs.
The reason is that log delivery makes the RW node write more log entries, which affects OLTP performance.
However, in the long run, it is regarded as almost impossible to synchronize heterogeneous storage with REDO logs~\cite{lee2017parallel}.
There are three challenges to this.
(1) REDO logs only record changes to physical pages in the row store and lack database-level or table-level information~\cite{supplog} 
(e.g., RO nodes do not know which table the page change corresponds to).
(2) Page changes caused by the row store itself rather than user DMLs are also included in REDO logs, such as B+tree splits/merges and page consolidations.
Column indexes cannot apply these logs, otherwise, inconsistencies may occur.
(3) REDO logs only include differences rather than complete updates to reduce log volume.

As shown in~\fig{fig:parallel}, PolarDB-IMCI addresses these challenges with two replay phases. 
The \textbf{Phase\#1} is to replay REDO logs to an in-memory copy of the row store in RO.
In this phase, \sysname{} captures the complete information to parse REDO logs into logical DML statements.
Then, the \textbf{Phase\#2} is to replay DML statements to column indexes.

The performance of replay is critical to our system.
To achieve high performance, several parallel replay mechanisms~\cite{f12020, butterstein2020replication, shen2021retrofitting, raman2013db2} are proposed in the literature.
These works either take parallel replay at session granularity or transaction granularity with the help of conflict-handling aids, 
such as locks or dependency graphs, or optimistic control.
Unlike these works, \sysname{} proposes a new replay approach, 2P-COFFER, to make both phases of parallel replay conflict-free.
In 2P-COFFER, the \textbf{Phase\#1} is page-grained, while the \textbf{Phase\#2} is row-grained to enable the concurrent modification of different pages/rows. 
Log entries that modify the same page/row but belong to different transactions are considered dependent and should be replayed sequentially. 
With 2P-COFFER, the replay throughput of RO nodes is much higher than the OLTP throughput of RW (\fig{fig:propagation:AP}).

\subsection{Phase\#1: Physical Log Parse}\label{subsec:logParse}
% when performing update propagation across formats,  
% it is hard to rely on physical logs since they are tightly coupled to the underlying storage.
% Therefore, in the long run, it is regraded as almost impossible to synchronizes heterogeneous storage with physical logs~\cite{hana}. 
% Generating logical DML statements from physical logs requires RO nodes to maintain row-format data in memory.
% does not generate additional logical logs to address this issue.
% The reason is that log delivery makes the RW node write more log entries, which affects OLTP performance.
% Instead, \sysname parses physical logs into logical DML statements that can be applied to column indexes.
% However, this is not a free lunch.
% Generating logical DML statements from physical logs requires RO nodes to maintain row-format data in memory.
% We take this trade-off as the performance of RW has a higher priority than RO in our scenario (described in ~\ref{}).
%Moreover, the in-memory row data of RO also lays the foundation for \sysname to support a hybrid query execution in the future.
% Before going into the exact parse steps, let's look at the structure of log entries. 
As shown in~\fig{fig:redo}, a REDO log entry of PolarDB contains multiple fields. 
For simplicity, we take the update operation as an example, and other sorts of operations are similar. 
\begin{itemize}
    \setlength{\itemsep}{0pt}
    \setlength{\parsep}{0pt}
    \setlength{\parskip}{0pt}
    \item TID is the transaction identifier that creates this entry. 
    \item LSN represents the order of this entry in the log. 
    \item PageID identifies which physical page the row updated by this entry belongs to. 
    The Offset field (SlotID) further determines where the updated row sits on the page. 
    \item Data field (Differential Log) contains the difference between the updated value and the original value. 
\end{itemize}

%Among them, PageID, slotID, and differential log are tightly tied to the physical location of the row-based store. 
%The differential log is even worse, part of it is expressed in terms of byte granularity (e.g., the third byte of new value).

%\sysname{} handles this problem with a physical log parser that parses physical logs into logical DML statements.
% It is straightforward to notice that the log structure of update operations does not contain a full new row. 
% As a result,
In the left part of~\fig{fig:parallel}, \textbf{Phase\#1} distributes REDO logs to different workers based on the PageID,
and each worker followed the LSN order to replay page changes to reproduce the DML details.
The dispatch process is similar to \textbf{Phase\#2} (\chref{subsec:logrepaly}) but at page granularity.
For an update-type log entry, the worker will generate a delete DML and an insert DML during replay as column indexes are updated out-place.
But the differential field of REDO logs may not contain PK information, which is required for deleting DMLs (find a row via the locator). 
Therefore, the worker gets the old row from PolarFS based on the PageID and offset field,  and uses the old rows to assemble a delete-type DML before applying for an entry.  
Then, the worker applies the differential field into the extracted rows to replay page changes,
and assemble the insert DML after applying.
To truly make up an operation into a logical DML, each operation must also be supplemented with its table schema.
Workers get table schema information by table IDs recorded on pages.

Furthermore, workers must identify log entries generated by the row store itself (e.g., B+tree splits).
To handle this, workers first check whether a log entry belongs to an active transaction by the TID.
If not, this entry is confirmed as not being generated by a user transaction.
If so, the worker further checks if the PK of this entry is repeatedly inserted in the active transaction (via a PK set).
Note that a duplicate PK insert is not a user DML.

Consequently, reusing REDO forces a replay of all page changes. As an optimization, \sysname{} let RO nodes maintain the buffer pool of the row store like RW to 
reduce the amount of data page reads.
In our practice, the computing capacity of \textbf{Phase\#1} is much greater than the log production capacity of RW.
On the one hand, RO nodes directly reproduce page changes without the overhead of redoing transactions, such as B+tree traversals. 
On the other hand, REDO logs under real workloads always act on hot pages so that the buffer pool has a hit rate close to 99\%. 
Although the buffer pool reduces the memory available for OLAP, we take this tradeoff because reducing the perturbation on OLTP through REDO logs is a higher priority in our scenario.

\subsection{Phase\#2: Logical DML Apply}\label{subsec:logrepaly}

REDO logs' LSN order ensures the fundamental prerequisite for log replay, which means changes in RO nodes can be made in the same order as RW.
\textbf{Phase\#1} breaks this order. 
Therefore, after the transformation, a background thread will sort DMLs according to the LSN of their associated log entries.
Then, the background thread inserts DMLs into transaction buffer units.

In \textbf{Phase\#2}, a dispatcher distributes a batch of transactions to multiple workers, 
performing modifications to column indexes in parallel.
The distribution is conducted row-by-row, and DML statements from a single transaction will be dispatched to multiple workers for replay.
For a DML statement, the dispatcher assigns a specified worker by taking a \textit{modulo} of the hash value of the row's primary key.
Therefore, DML statements that modify the same row are assigned to the same worker in the commit order, even if they belong to different transactions.
%Ensuring different modifications to the same row are delivered to the same worker in order guarantees consistency.
The dispatcher processes each transaction in the commit order, 
ensuring that different modifications to the same row are delivered to the same worker in order, which guarantees consistency.
% In actuality, the commit order of a transaction is indicated by the LSN of its commit log entry.
Each worker follows the steps described in~\chref{sec:column:dml} to replay each DML statement in order, and changes will be committed to column indexes in batch.
% As a result, there is no conflict in the parallel replay of transactions,  increasing the write throughput of \sysname.

The right part of \fig{fig:parallel} illustrates how two workers ($W_1$ and $W_2$) can replay two transactions ($T_1$ and $T_2$) simultaneously.
$T_1$ Insert ($1$, ``$A$'') and Insert ($2$, ``$D$'') respectively. $T_2$ Update (2, ``$B$'') and Insert (3, ``$C$'').
Insert ($2$, ``$D$'') and Update ($2$, ``$B$'') are assigned to $W_2$ with the commit order of $T_1$ and $T_2$. 
$W_1$ executes these two DMLs in sequence without concurrent conflicts.

\subsection{Handle Large Transactions}

So far, we have presented the update propagation of \sysname{}, but there is one more issue.
As stated in~\ref{subsec:CALS}, CALS prefetches log entries from PolarFS into transaction buffers.
Therefore, if a transaction comprises too many operations, its transaction buffer unit may consume a huge memory.

To avoid excessive memory consumption, \sysname{} pre-commits large transactions: 
DML statements in a transaction buffer unit are pre-committed when their number reaches a given threshold.
The basic idea behind pre-commitment is to write updates to Partial Packs with invalid insert and delete VIDs, 
rendering the updates temporarily invisible. 
The specific steps of pre-commitment are as follows.
First, request a continuous RID for all rows in the current transaction buffer, and save this RID range. 
It is important to note that the global RID locator cannot yet be changed during the pre-commit phase 
to avoid the exposure of uncommitted transactions.
Thus, \sysname{} creates a temporary RID locator instead of updating the RID global locator to cache new PK-to-RID mappings. 
Then, \sysname{} writes the updates to Partial Packs while setting the insert and delete VIDs as invalid to make them invisible. 
Finally, \sysname{} frees the memory used by the transaction buffer unit.

When the large transaction commits, \sysname{} merges the temporary RID locator into the global RID locator 
and rectifies the invalid VIDs (in the saved RID range) with the transaction commit sequence number. Otherwise, if the large transaction aborts, the temporary locator will be cleaned out.
Pre-commit rows remaining in Partial Packs are invalid and will be eliminated later by compaction threads in the background.

\section{Analytical Processing}\label{sec:olap}
% This section describes how PolarDB IMCI identifies and processes analytical queries.
% The whole process is divided into three parts: routing, translation, and execution, which we will introduce one by one.
% Last but not least, we also cover how analytic queries achieve strong consistency.

\subsection{Transparent Query Routing}\label{sec:olap:routing}
In \sysname{}, queries can be executed on different nodes and different execution engines via a cost-based routing protocol.
The routing process is completely transparent to applications and users and has a two-levels policy: inter-node routing and intra-node routing.
Inter-node routing implements read/write flow splitting (with load balance) through the proxy layer, 
while intra-node routing provides a dynamic selection of data access paths and execution engines (either row-based or column-based) through the optimizer.

\noindent\textbf{Inter-node Routing.}
\sysname{}'s proxy provides a unified SQL interface for all application requests (both OLTP and OLAP).
When requests come in, the proxy directs read/write requests (e.g., transactions) to the RW node and directs read-only queries 
(e.g., analytical queries) to RO nodes via a rough syntax parser. If multiple RO nodes are deployed, the proxy will balance the 
traffic based on the number of active sessions.

\begin{figure}  
    \centering
    \includegraphics[width=\columnwidth]{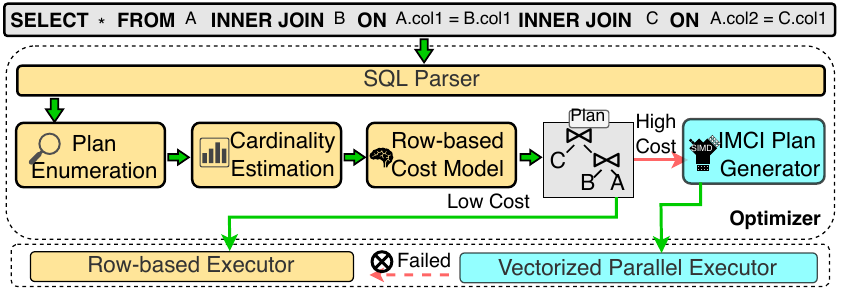}
    \vspace{-2em}
    \caption{\small The workflow of \sysname{}'s optimizer.}\label{fig:AP:opt}
    \vspace{-2em}
\end{figure}

\noindent\textbf{Intra-node Routing.} As shown in \fig{fig:AP:opt}, \sysname{} implements two execution engines within each RO node.
A row-based execution engine for point queries and a column-based execution engine for analytical queries. 
The optimizer of \sysname{} selects the appropriate execution engine for each query based on row-based cost estimation.
By assuming that all queries can preferentially run in the row-based execution engine (i.e., low cost), the optimizer generates a row-oriented execution plan first. 
If the estimated cost of the row-oriented plan exceeds a threshold (i.e., high cost), a column-oriented plan will be generated and used over the column-based engine. 
%We implement such an intra-node routing policy based on row-based cost estimation instead of implementing a new hybrid cost model, 
%since row-based cost estimation is much more robust and well-studied.
We leave the exciting development of a new row-column hybrid cost model and hybrid execution plan as our future work. 
% The estimated cost is accurate most of the time in that it takes into account multiple factors, including IO cost, CPU cost and data volume.
% PolarDB IMCI allows customers to force the executor selection when the estimation is inaccurate.

% The issue of intra-node routing is essentially a result of PolarDB using two executors.
% Currently, PolarDB support for row-column hybrid execution is under development. 
% In the future a single executor will be able to juggle both point queries and table scans, and then inter-node routing will no longer be required.

\subsection{IMCI Plan Generation}\label{sec:olap:plangen}

% Unlike many multi-node HTAP solutions~\cite{} that employ separated executors, 
% column-oriented execution plans in PolarDB IMCI are not entirely top-down constructed.
Instead of top-down constructing a column-oriented execution plan, 
\sysname{} transforms it from the row-oriented one.
  %Due to the differences between execution engines, column-oriented plans benefit less from following the join order of row plans.
  %Thus, the optimizer further optimizes the join order after constructing a column-oriented plan.
  \sysname{} uses DPhyp~\cite{MoerkotteN08} as the join ordering algorithm and collects statistics through random sampling~\cite{haas1998estimating,ChaudhuriMN98,ChaudhuriDS04}. 
  %, which can efficiently handle various types of joins, including outer-joins and anti-joins.
  %To provide accurate cardinality estimation for the optimizer, \sysname{} collects statistics through random sampling. 
  %Sampling tasks will be performed periodically in the background.  
  %Furthermore, \sysname{} adaptively adopts various sampling methods~\cite{haas1998estimating,ChaudhuriMN98,ChaudhuriDS04} to make statistics efficient and accurate.
 The transform workflow is shown in~\fig{fig:AP:opt}. By doing so, column-oriented plans can preserve all behavioral characteristics.

For instance, in \sysname{}, implicit type casts of a column-oriented plan are always consistent with the row-oriented plan.
During the plan generation,  \sysname{} transforms the original expressions into a vectorized execution format to exploit SIMD instructions.
This transformation is handled inside the expression objects (e.g.,~\textit{Item} class in \textit{MySQL}) and strictly follows up on original implicit type casts.
Another instance is that column-oriented plans reuse error codes and messages from row-oriented plans.
It is challenging to align errors across different execution engines.
In \sysname{}, Column-oriented plans can retain static error detection directly from row-oriented ones to avoid this issue.
For run-time errors, \sysname{} will fall back the execution to be row-oriented.
% Instead of aggressively using vectorized execution engine, \sysname{} adoptes a bottom line strategy.
%\sysname{} implements such a fallback policy by maintaining a white-list. 
%The white-list identifies whether a functionality is supported by the column-based execution engine, including expressions, syntax, etc.
%After the SQL parse phase, \sysname{} examines each item in execution plan against the white-list.
%If not supported, the SQL will not be scheduled to the column-based engine and will fall back to the row-based execution.
As a result, \sysname{} achieves strong compatibility with the existing framework of \textit{MySQL}.

\subsection{Execution Engine}\label{sec:olap:exe}
 
To obtain advanced OLAP performance, \sysname{} designs a new high-performance analytical execution engine (i.e., column-based engine).
Drawing on the knowledge of in-memory columnar databases~\cite{LeisBK014Morsel,BonczZN05MonetDB,FarberMLGMRD12HANA}, the analytical engine incorporates numerous state-of-the-art technologies, 
including a pipeline execution model, a set of well-optimized parallel operators, and a vectorized expression evaluation framework.
\begin{itemize}[leftmargin=*]
  %  \vspace{-3pt}
  \setlength{\itemsep}{0pt}
  \setlength{\parsep}{0pt}
  \setlength{\parskip}{0pt}

\item \textbf{Pipeline Execution.}
The execution tree of a vectorized execution plan is decomposed into multiple linear paths called pipelines. 
In a pipeline, a non-blocking operator (e.g., Filter, Join Probe) processes one batch at a time 
instead of all data, and then passes the intermediate result to the next operator.
Pipeline execution brings several advantages: (1). a batch of data that streams through multiple operators is always cached;
(2). intermediate results are reduced to minimize the memory footprint.
\item \textbf{Parallel Operators.}
To parallelize each pipeline, all operators in the analytical engine support parallel execution. 
For example, TableScan can concurrently fetch Data Packs in a non-interleaved manner, 
and the analytical engine implements Join as a lock-free partition Join~\cite{BalkesenTAO13ParJoin}.
Furthermore, to reduce cache misses, blocking operators use carefully designed data structures (e.g.,cache-friendly hash tables~\cite{BarberLPRSACLS14CHT}) 
and software prefetching~\cite{ChenAGM04Prefetch} as much as possible.
Besides, all blocking operators have an optimized spill-to-disk version 
to handle out-of-memory crises, such as dynamic hybrid hash Join~\cite{JahangiriCF22DJoin}.

\item \textbf{Expression Evaluation.} 
When a batch of data is cached, the performance bottleneck is switched from memory access to CPU computation.
SIMD instructions, sometimes known as vectorized instructions, such as AVX-512, are powerful for accelerating CPU computation.
Thus, an expression evaluation framework~\cite{ROVEC} is decoupled from operators to serve compute-intensive modules in a vectorized manner (i.e., using SIMD).

\end{itemize}

\subsection{Strong Consistency}\label{sec:olap:consistency}

Since \sysname{} uses an asynchronous replication mechanism, analytical queries may observe stale data.
For example, an analytical query may not read the updates that have already been committed in the RW node.
However, it is possible for \sysname{} to achieve multiple consistency levels through the proxy layer to meet the
requirement of applications, including strong consistency.

The proxy keeps track of the RW node's written LSN and all RO nodes' applied LSN.  
The written LSN and applied LSN indicate the transaction commit point for RW and RO, respectively.
Transactions before the written LSN were committed on the RW node.
Likewise, any log entries before the applied LSN are guaranteed to have been replayed by the RO node.
The proxy may only route queries to the RO nodes whose applied LSN is not less than the written LSN to meet the requirements of strong consistency.
% If none of the RO nodes meet the requirement, the query waits until at least one of them does or times out.
% When times out, the proxy routes the query to the RW node, or returns an error.

\section{Resource Elasticity}\label{sec:fast_startup}

One of the core design concepts behind \sysname{} is to 
realize on-demand node provisioning with a storage-computation separation architecture. In this section, we dive into the node scale-out mechanism in \sysname. 

Like most in-memory database systems~\cite{FarberMLGMRD12HANA,oracle2015}, \sysname periodically stores column indexes in shared storage as checkpoints 
to provide fast recovery after a system crash.
More specifically, in \sysname, new scale-out RO nodes can quickly construct their memory structures with checkpoints.
In our implementation, the roles of RO nodes are divided into one leader and multiple followers.
A leader is in charge of issuing checkpoints, while followers maintain their own memory structures, and leverage the checkpoints for fast recovery. The role assignment is centrally controlled by RW. 
% When \sysname initially adds an RO node, RW designates this RO node as the leader.
When start-up, RW designates the first RO node in the cluster as an RO leader, and other RO nodes are followers.
If the leader crashes, RW will re-designate one of the followers to be the new leader.

% During a checkpoint, it is important not to block log replay for too long;
% otherwise, replay throughput would suffer greatly.

%\begin{figure*}[t]  
%  \centering
% \setlength{\belowcaptionskip}{-5pt} 
  % \setlength{\abovecaptionskip}{6pt} 
%  \includegraphics[width=2\columnwidth]{eval-figs/OLAP_execution_time.pdf}
%   \vspace{-1.2em}
%  \caption{\small Comparison of \sysname, PolarDB, and ClickHouse on TPC-H (100GB). All systems used 32 threads for intra-query parallelism. } \label{fig:AP:perf:small}
%  \vspace{-.6em}
%\end{figure*}

%\begin{figure*}[t]  
%  \centering
% \setlength{\belowcaptionskip}{-5pt} 
  % \setlength{\abovecaptionskip}{6pt} 
%  \includegraphics[width=2\columnwidth]{eval-figs/OLAP_execution_time_1TB.pdf}
%   \vspace{-1.2em}
%  \caption{\small Comparison of \sysname, PolarDB, and ClickHouse on TPC-H (1TB). All systems used 32 threads for intra-query parallelism. } \label{fig:AP:perf:large}
%  \vspace{-.6em}
%\end{figure*}

\begin{figure*}
  \centering
%  \subfigure[100$GB$]{
%    \begin{minipage}[b]{0.5\textwidth}
%    \includegraphics[width=1\textwidth]{eval-figs/OLAP_execution_time.pdf}
%    \end{minipage}
%  }
  \begin{subfigure}[t]{0.98\textwidth}
		\centering
		\includegraphics[width=\columnwidth]{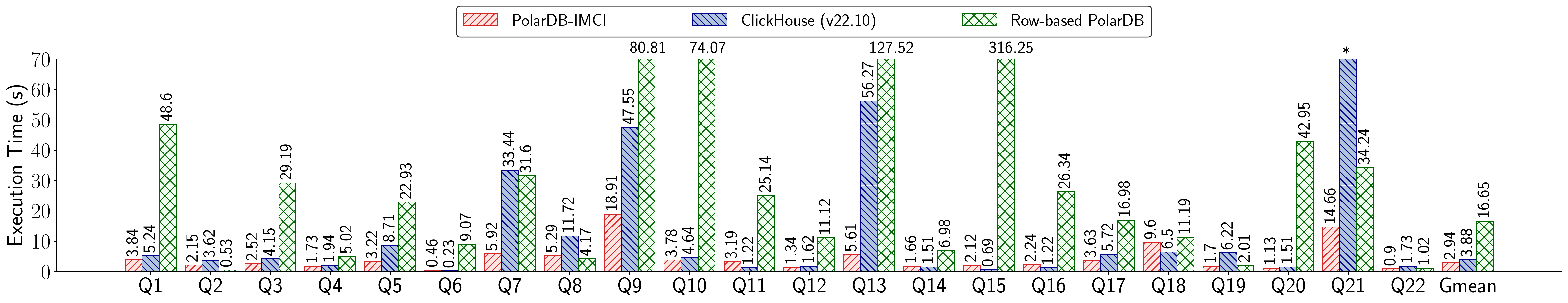}
		\vspace{-4ex}
		\subcaption{$100\,GB$}\label{fig:AP:perf:small}
		% \vspace{1em}
  \end{subfigure}
  
  \begin{subfigure}[t]{0.98\textwidth}
		\centering
		\includegraphics[width=\columnwidth]{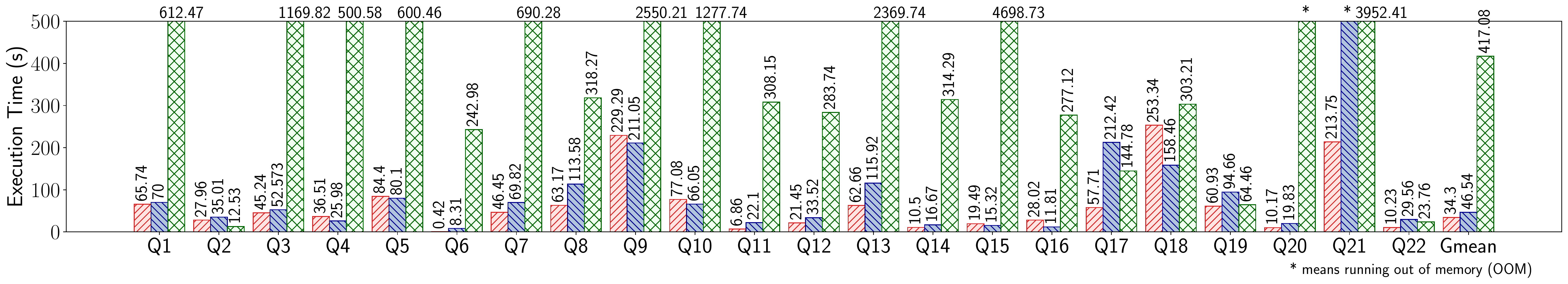}
		\vspace{-6ex}
		\subcaption{$1\,TB$}\label{fig:AP:perf:large}
		% \vspace{1em}
  \end{subfigure}
  \vspace{-1.2em}
  \caption{\small Comparison of \sysname, PolarDB, and ClickHouse on TPC-H. All systems used 32 threads for intra-query parallelism.} \label{fig:AP:perf}
  %\vspace{-1.2em}
\end{figure*}

To take a checkpoint, the leader identifies the latest committed transaction sequence number as the Checkpoint Sequence Number (CSN). The transactions committed after the CSN are excluded in the checkpoint to enable a consistent data view between RO nodes.
A major challenge is to ensure that the checkpointing tasks never stall the foreground log replay. 
However, checkpointing tasks may be stained when the log replay is in progress. 
Recall that there are three important in-memory structures (the RID locator, Packs, and VID maps) in RO nodes, and all of them should be coordinated with checkpoints. 
Addressing the challenge, \sysname{} handles each of them by the following steps.
\begin{itemize}[leftmargin=*]
  %\vspace{-1em}
  \setlength{\itemsep}{0pt}
  \setlength{\parsep}{0pt}
  \setlength{\parskip}{0pt}
  \item  
  % Thanks to our design on data Packs, updates to Packs and Partial Packs never stain the checkpoint.
  % On the one hand, Packs are immutable. On the other hand, updates to Partial Packs are append-only.
  Packs in \sysname{} are append-only and immutable, which means the persistence timing of Packs 
  is independent of checkpoints.
  Hence, Packs on the leader are written into PolarFS as soon as they are created. Visibility is controlled by VID maps.
  \item  VID maps require a more careful design. Firstly, \sysname{} generates a copy of VID maps on the leader 
  and parallelly checks all elements in the copy. If VIDs exceed the CSN, the elements in VID maps will be marked as invalid.
  Then, the visibility controlled by VID maps is aligned with the CSN and the copy can be persisted into PolarFS.
  Note that, when start-up, the replayed transactions will allocate new slots of column index in Partial Packs for insertions, or remove the valid marks for deletions.
% Transactions that exceed the CSN will be replayed after the system restart.
% Log entries may be replayed repeatedly if the exceeded VIDs are still valid.
  \item RID locator splits a new immutable copy for checkpoints tasks by functional data structures~\cite{driscoll1989making}.
% Each time issuing a checkpoint, the locator creates a new version. 
Therefore, Subsequent transactions will not stain the checkpoint. 
Meanwhile, to prevent active transactions from leaving residues on the old view, checkpoints are only triggered when the MemTable of LSMTree is filled.
% Moreover, the locator also applies approaches (e.g., functional data structures~\cite{}) optimized for multi-versions to reduce the version changeover cost.
\end{itemize}

When adding a new RO node, \sysname{} first checks whether there is an available checkpoint of column indexes in PolarFS.
If so, it loads the checkpoint and performs fast recovery; otherwise, it rebuilds column indexes from the row store.
After that, the RO node replays the log entries after the checkpoint to catch up with the RO leader. 
% During catching up, the RO node is able to serve queries with poor freshness. 
% The poor freshness lasts only a short time.
The experiments in~\chref{sec:eval:elasticity} show that scaling out a RO node takes tens of seconds.

% If recovering data from a checkpoint, RO nodes load the locator and  VID maps into memory first. 
% Regarding Packs, RO nodes use a lazy loading way. 
% Only Packs accessed by queries are loaded into memory to reduce startup time.
% Noting that lagging Pack loads do not bring up future checkpoints as Packs are immutable.

% \clearpage

\section{EVALUATION} \label{sec:eval}

\subsection{Evaluation Setup}\label{sec:eval:setup}
\noindent\textbf{Configurations.} 
The experimental evaluation was carried out on a \sysname{} cluster (mmx8.4xlarge) in the Alibaba Cloud platform.
Except for the scale-out experiment, we used two computation nodes, one read/write (RW) node and one read-only (RO) node. 
The instances are attached to a PolarFS volume which can provide nearly unlimited capacity.
We used one ECS server (c7.8xlarge) on Alibaba Cloud as HTAP clients to issue SQL requests. The detailed configurations can be found in~\tab{tab:cluster}. 
% The scale-out experiment was conducted by adding RO nodes, thus consuming more than two nodes. 

\noindent\textbf{Benchmarks.} To emulate diverse application scenarios and analyze the performance of \sysname{} systematically, we used three
well-studied and widely-used benchmarks. 

%TPC-H~\cite{tpch} is adopted to evaluate the performance of \sysname{} in executing analytical queries. We used 100 $GB$ and 1 $TB$ of data volume,
%and reported the running time of each query. 
%We also reported the geometric mean of all 22 queries, as suggested in the TPC-H official document. 

To evaluate  \sysname{}'s performance in executing analytical queries, we adopted TPC-H with 100 $GB$ and 1 $TB$ data volume. 
We reported the running time of each query and the geometric mean of all 22 queries, as suggested in the TPC-H official document. 

We used CH-benCHmarks~\cite{cole2011mixed} to evaluate \sysname{}'s performance under hybrid workloads. It integrates TPC-H queries into TPC-C~\cite{tpcc} with a unified data schema. 
We reported the OLTP and OLAP throughput and studied the performance isolation property with a scale fact (i.e., the number of data warehouses) = 100. 

To provide a more in-depth analysis of \sysname{}'s micro component, sysbench~\cite{sysbench} is used for pressure tests with diverse workload patterns.
We set insert-only and write-only (update) workloads with Zipfian distribution. The database contains 100 tables using 64-bit integers as
primary keys and 188 bytes per record.

We ran each experiment 10 times and reported the average number. 
Results were collected in the middle of each experiment to avoid the disturbance caused by system start-up and cool-down. 

Our evaluation focused on the following questions:
\setlist[itemize]{leftmargin=7.5mm}
\begin{itemize}
	\item[\chref{sec:eval:overall}] What is the overall performance of \sysname{}?
    \item[\chref{sec:eval:freshness}] Can \sysname{} achieve high data freshness?
    \item[\chref{sec:eval:isolation}] How does \sysname{} handle update propagation?
    \item[\chref{sec:eval:elasticity}] Can \sysname{} achieve high resource elasticity when OLAP workloads increase? 
    \item[\chref{sec:eval:production}] How does \sysname{} benefit real-world applications in production deployment?
    % \item[\chref{sec:eval:discussion}] What have we learned from \sysname?
\end{itemize}

\subsection{Overall Performance}\label{sec:eval:overall}
% {\color{blue} \noindent\textbf{OLTP-only workloads} } Argument: OLTP performance is advanced (compare to MySQL) and comparable to PolarDB without IMCI. \fig{fig:TP}

\noindent\textbf{OLAP-only workloads.} Achieving advanced OLAP performance (i.e., \textbf{G\#2}) in an HTAP system 
is one of the foremost motivations of \sysname{}. 
In this evaluation, we compared the TPC-H query execution time of \sysname{}'s column execution engine,  
its row execution engine (referred to as row-based PolarDB),
and ClickHouse (an advanced OLAP system).
For an apple-to-apple comparison, we built secondary indexes for each column in row-based PolarDB to maximize its performance.
Currently, ClickHouse does not offer enough support for join reordering~\cite{ckjoin}.
To further evaluate the performance of execution engines, in the 1 $TB$ experiment, we manually adjusted the join order of ClickHouse to the same as \sysname{}.
Queries are executed one by one, and all systems used 32 threads for intra-query parallelism.

\begin{table}[t]
	%\vspace{-5pt}
	\small 
	% \setlength\tabcolsep{7.5pt} 
	% \begin{tabular}{l|l}
	% 	\hline
	% HTAP Clients; RW Node; RO Node with IMCI & 1 server for each \\ \hline                                               
	% \end{tabular}
	\caption{Configurations of our evaluation.}\label{tab:cluster}
	\vspace{-5pt}
	\setlength\tabcolsep{3pt} 
	\begin{tabular}{l|l}
		\hline \hline
	RW/RO Node    & \begin{tabular}[c]{@{}l@{}} 32 vCPU, 1 NUMA node\\ 256 $GB$ DRAM \end{tabular} \\ \hline
%	CPU     & \begin{tabular}[c]{@{}l@{}} 32 vCPUIntel(R) Xeon(R) Platinum 8269CY @ 2.50GHz \\ (1 NUMA node, 32 cores) \end{tabular} \\ \hline
	Client   & \begin{tabular}[c]{@{}l@{}} 32 vCPU \\ 64 $GB$ DRAM \end{tabular} \\ \hline
%  Memory  & 256GB                                                                  \\ \hline
%	I/O bandwidth & 24Gbit/s                                                        \\ \hline 
%	Cache   & L1d \& L1i cache: 32K; L2 cache: 1024K; L3 cache: 36608K    \\ \hline
	OS      & Alibaba Group Enterprise Linux Server release 7.2        \\ \hline
	Network & 10Gbit/s Bandwidth                                               \\ \hline 
	PolarFS & \begin{tabular} [c]{@{}l@{}} 288000 IOPS (RandRead 16KB) \\ 18000 IOPS (SeqWrite 128KB)  \end{tabular} \\ \hline   

\end{tabular}
\vspace{-1.0em}
\end{table}

\fig{fig:AP:perf} shows the results.
With 100 $GB$ data, \sysname{} achieved $\times$5.56 speed-ups (in geometric mean) compared to row-based PolarDB, and up to $\times$149.12 speed-ups for scan-intensive queries (e.g., $Q10$, $Q15$). With 1 $TB$ data, the speed-ups are $\times$12.15 in geometric mean.
The performance gain came from two folds: first, \sysname{} serves scan operations on column granularity, which minimizes read amplification caused by full table scan (\chref{sec:column}); 
second, \sysname{} implements parallel operators, batch iteration, and SIMD optimizations to speed query processing on the large data volume (\chref{sec:olap}). 
One may find $Q2$ interesting: \sysname{} underperformed on such queries. This was because the selectivity of $Q2$ was low, 
and indexes built in row-based PolarDB were more efficient to handle point queries. However, thanks to our optimizer (\chref{sec:olap:routing}), 
in practice, \sysname{} can automatically route such queries to their desirable execution engine.
Compared to ClickHouse, \sysname{} outperformed or was competitive with it on most queries. 
Overall, \sysname{} achieved $\times$1.32 speed-ups on 100 $GB$ data and $\times$1.35 speed-ups on 1 $TB$ data.
\sysname{} incurred longer execution time on a small specific set of queries (e.g., $Q11$, $Q18$). 
%This is because \sysname{} implemented most technical optimizations in ClickHouse. And different implementations of operators and query optimizations differ in the OLAP performance for specific queries.
% To the best of our knowledge, this is caused by different implementation details in operators and memory management.
% For instance, ClickHouse implemented streaming aggregator, and it outperformed hash aggregator in $Q18$ when the raw data was ordered.
In summary, \sysname{}'s OLAP performance is much better than row-based PolarDB and is comparable to ClickHouse.

\begin{figure}[t]
	\centering
	\begin{subfigure}[t]{0.49\columnwidth}
		\centering
		\includegraphics[width=\columnwidth]{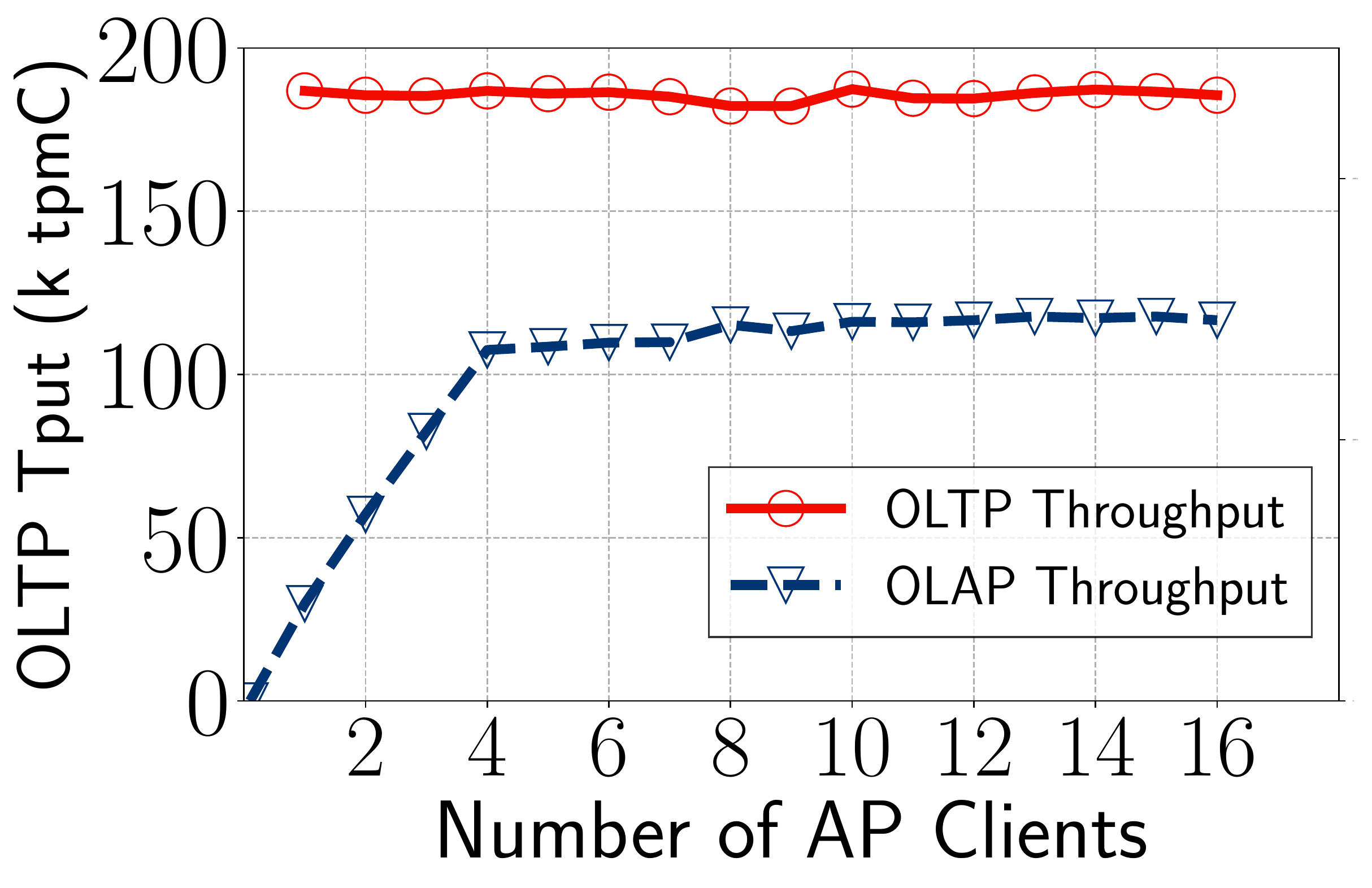}
		\vspace{-4ex}
		\subcaption{Perf. isolation of OLTP}\label{fig:isolation:TP}
		% \vspace{1em} 
    \end{subfigure}
	\begin{subfigure}[t]{0.47\columnwidth}
		\centering
		\includegraphics[width=0.98\columnwidth]{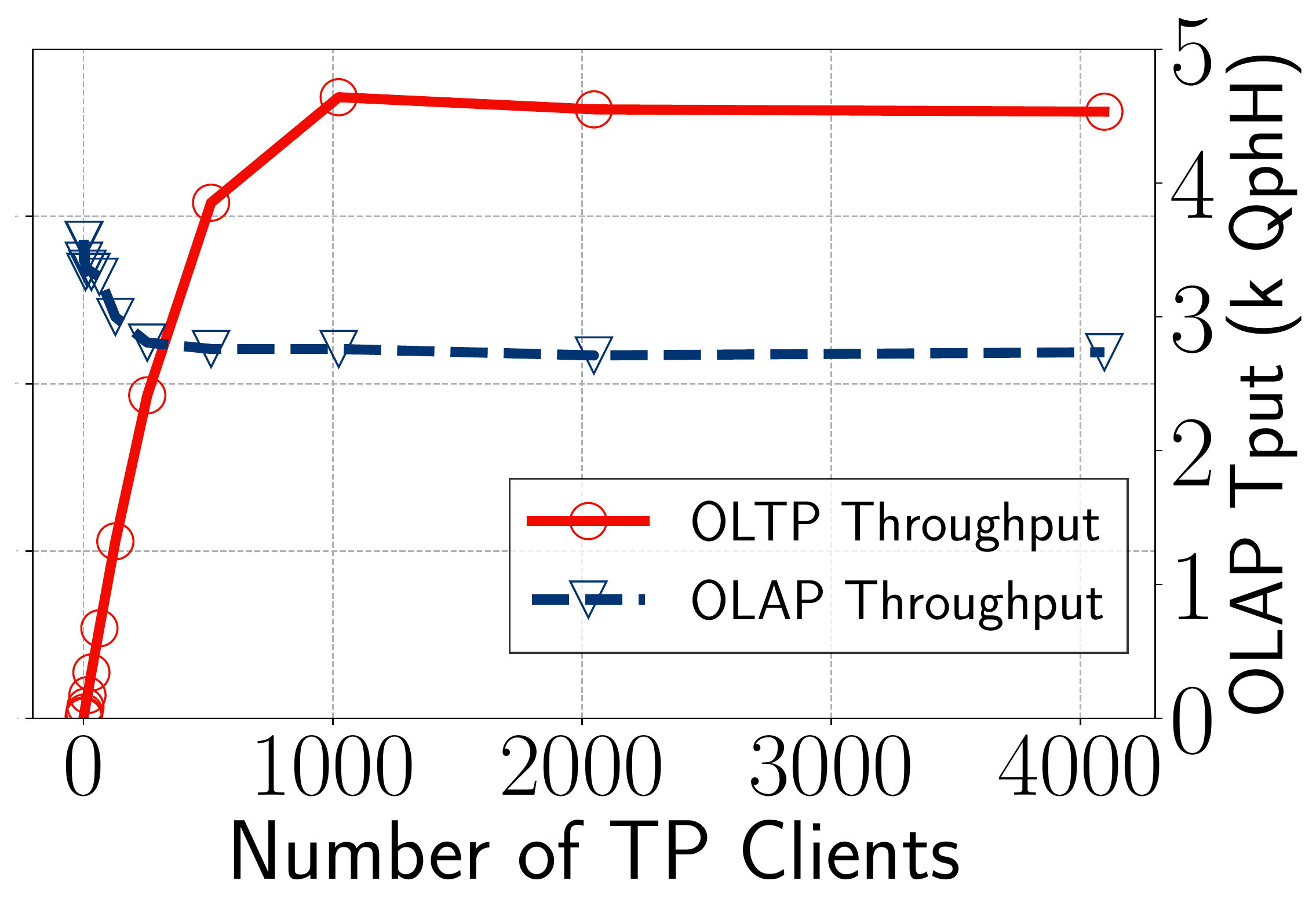}
		\vspace{-1.5ex}
		% \vspace{1pt}
		\subcaption{Perf. isolation of OLAP}\label{fig:isolation:AP} 
    \end{subfigure}
     \vspace{-1.2em}
\caption{\small Isolated OLTP and OLAP Performance of \sysname{} on CH-benCHmark Workloads. These two sub-figures share the same y-axis.}\label{fig:isolation} 
\vspace{-0.5em}
\end{figure}

\begin{figure}[t]  
    \centering
    \includegraphics[width=1\columnwidth]{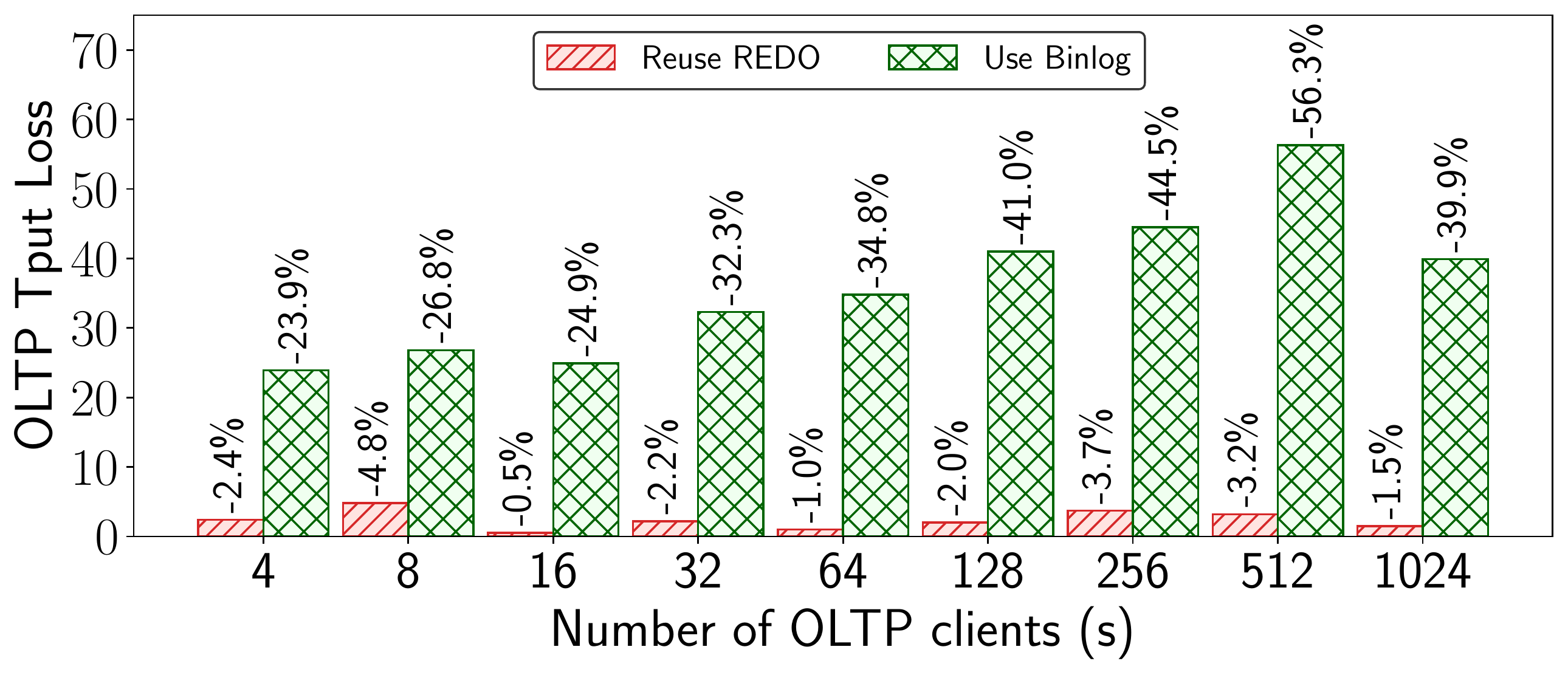}
    \vspace{-1.5em}
    \caption{\small Effectiveness of updates propagation methods. The Loss is calculated by comparing PolarDB without IMCI.} \label{fig:propagation:TP}
	\vspace{-1em}
\end{figure}

% \begin{figure}[t]  
%     \centering
%     \includegraphics[width=0.5\columnwidth]{eval-figs/Propagation_AP.pdf}
%     \vspace{-2em}
%     \caption{\small Effectiveness of IMCI's Parallel Replay on Sysbench.} \label{fig:propagation:AP}
% \end{figure}

\noindent\textbf{HTAP workloads.} We test \sysname{}'s performance on hybrid workloads with CH-benCHmarks (\chref{sec:eval:setup}).
The results in \fig{fig:isolation} show that PolarDB-IMCI has effective resource isolation.
Following the standard~\cite{shen2021retrofitting}, we evaluate \sysname{} in two rounds. First, we used 512 OLTP clients to saturate OLTP throughput (i.e., tune the
number of clients to use 80\% of CPU resources), and increased OLAP clients to issue analytical queries. As \fig{fig:isolation:TP} demonstrated, \sysname{} can 
perform at most 186890 tpmC (TPC-C NewOrder transactions per minute) and 2916 QphH (TPC-H query per hour) simultaneously, and the performance 
throughput degradation of OLTP is low (less than 1\%).
Second, we changed the roles of OLTP and OLAP, and let OLTP workloads increase after OLAP throughput was saturated. 
\sysname{} indeed incurred a little throughput degradation on OLAP throughput (\textless 20\%). 
We conclude this degradation for two reasons:  
(1). OLTP workloads enlarged the table size of some tables, thus higher OLTP throughput may degrade more OLAP performance; 
(2). the number of invalid rows in Packs increased with higher OLTP throughput.
It validates our design choice of building column indexes on separated RO nodes.

We omit the comparison between \sysname{} and other transactional databases because the OLTP performance of \sysname{} strictly follows the performance of PolarDB~\cite{cao2021polardb, Cao2018}. \sysname{} achieved good performance isolation between workloads (see \fig{fig:isolation}) and the overhead of enabling IMCI is low (see \fig{fig:propagation:TP}).

\subsection{Performance Perturbation}\label{sec:eval:isolation}
Then, we examine how the update propagation affects PolarDB's OLTP performance.
Recall that minimal perturbation on OLTP (i.e., \textbf{G\#3}) is pivotal to our consumers' experience.
%\noindent\textbf{Effect of Updates Propagation on OLTP.} 
We design this experiment based on the sysbench insert-only workload,
%All secondary indexes were removed (disabled) to minimize resource contention.
and calculated the throughput loss by comparing the throughput of 
candidate methods to the original throughput without IMCI (i.e., PolarDB with only row-based read-only replica).
We started the experiment with an empty table and warmed up for 10 seconds.
\fig{fig:propagation:TP} shows the results. Compared to using Binlog, 
\sysname's updates propagation methods (i.e., reusing REDO log) caused minimal performance perturbation to OLTP. 
The overhead of using Binlog was significantly higher because Binlog incurred additional fsyncs and more log IO. 
% Applying Binlog in REDO also incurred additional unneglectable overhead, since the overhead on spawned write operations. 
One may consider the drawback of reusing REDO is that \sysname{} has to parse physical logs to logical logs. However, it does not cause 
a bottleneck in log replay, as validated in our experiment (\chref{sec:eval:freshness}).

\begin{figure}[t]
	\centering
	\begin{minipage}[t]{0.49\columnwidth}
		\centering
		\includegraphics[width=\columnwidth]{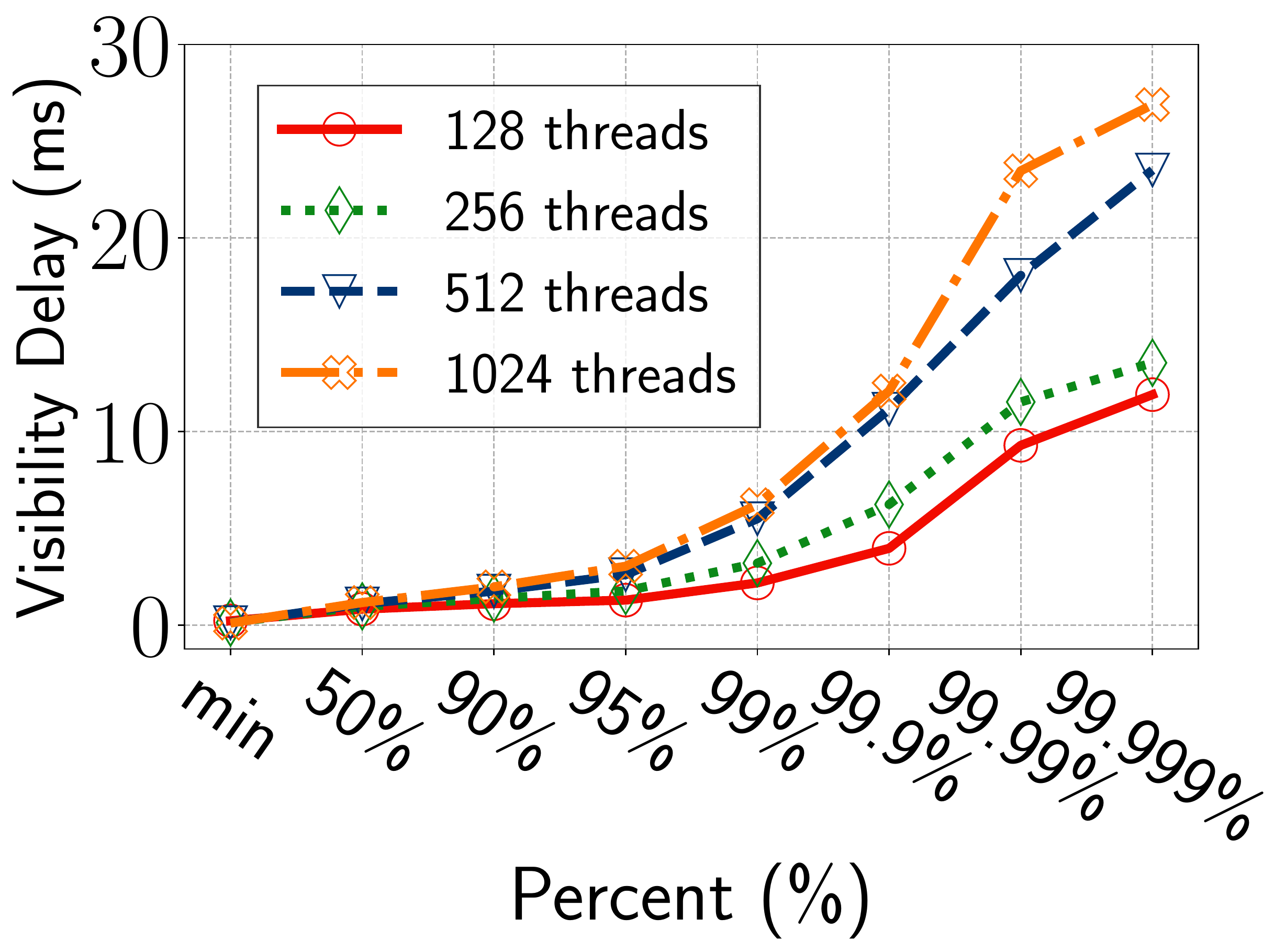}
		\vspace{-6ex}
		\caption{\small VD on TPC-C.}\label{fig:freshness} 
		% \vspace{1pt}
    \end{minipage}
	\begin{minipage}[t]{0.48\columnwidth}
		\centering
		\includegraphics[width=\columnwidth]{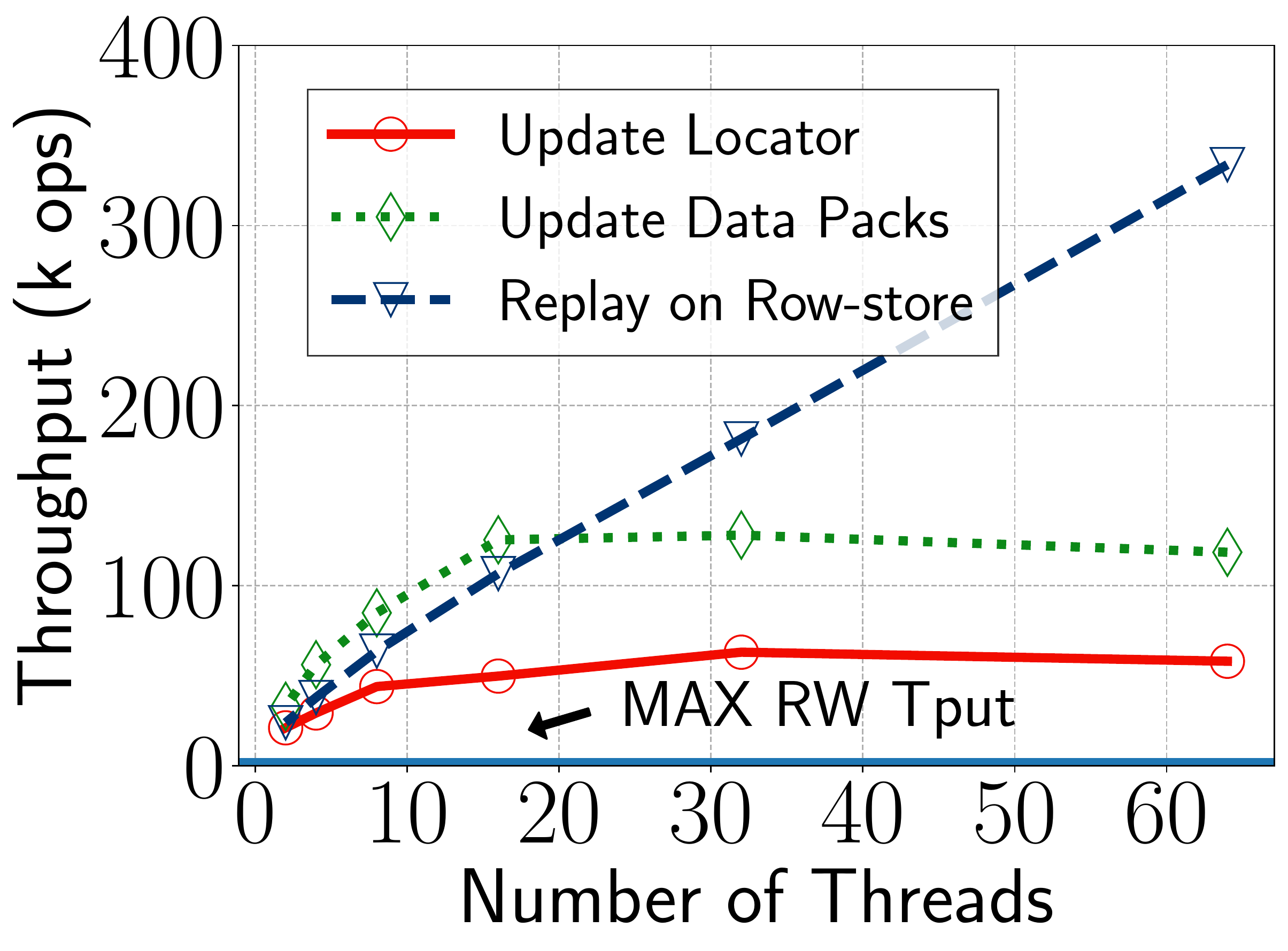}
		\vspace{-6ex}
		\caption{\small Replay Performance.} \label{fig:propagation:AP}
		% \vspace{1em}
    \end{minipage}
% \caption{Timeliness and visible delay evaluation of \xxx and three HTAP systems on interactive cross-site queries (HTAPBench-E) and skewed workload (Microbench-zipf).}
\vspace{-0.5em}
\end{figure}

% \begin{figure}  
%     \centering
%     \includegraphics[width=1\columnwidth]{eval-figs/scale.pdf}
%     \vspace{-2em}
%     \caption{\small Effectness of Scale-out and scaleup (add RO nodes with IMCI or upgrade AP instances) .} \label{fig:scaleout}
%     \vspace{-1.6em}
% \end{figure}

%\begin{figure}[t]
%	\centering
%		\centering
%		\includegraphics[width=\columnwidth]{eval-figs/scale-up.pdf}
		% \vspace{-4ex}
		% \vspace{1pt}
%     \vspace{-2.5em}
%\caption{\small Resource Elasticity on TPC-H (Scale Up)}\label{fig:AP:scaleup}
%    \vspace{-1.0em}
%\end{figure}

% \begin{figure}[t]
% 	\centering
% 	\begin{minipage}[t]{0.48\columnwidth}
% 		\centering
% 		\includegraphics[width=\columnwidth]{eval-figs/Propagation_AP.pdf}
% 		\vspace{-4ex}
% 		\caption{\small Parallel Replay} \label{fig:propagation:AP}
% 		% \vspace{1em}
%     \end{minipage}
% 	\begin{minipage}[t]{0.5\columnwidth}
% 		\centering
% 		\includegraphics[width=0.95\columnwidth]{eval-figs/scale-up.pdf}
% 		\vspace{-1.5ex}
% 		\caption{\small Scale-up Elasticity}\label{fig:AP:scaleup}
% 		% \vspace{1pt}
%     \end{minipage}
% % \caption{Timeliness and visible delay evaluation of \xxx and three HTAP systems on interactive cross-site queries (HTAPBench-E) and skewed workload (Microbench-zipf).}
% \vspace{-1.5em}
% \end{figure}

\begin{figure}[]
	\centering
		\centering
		\includegraphics[width=\columnwidth]{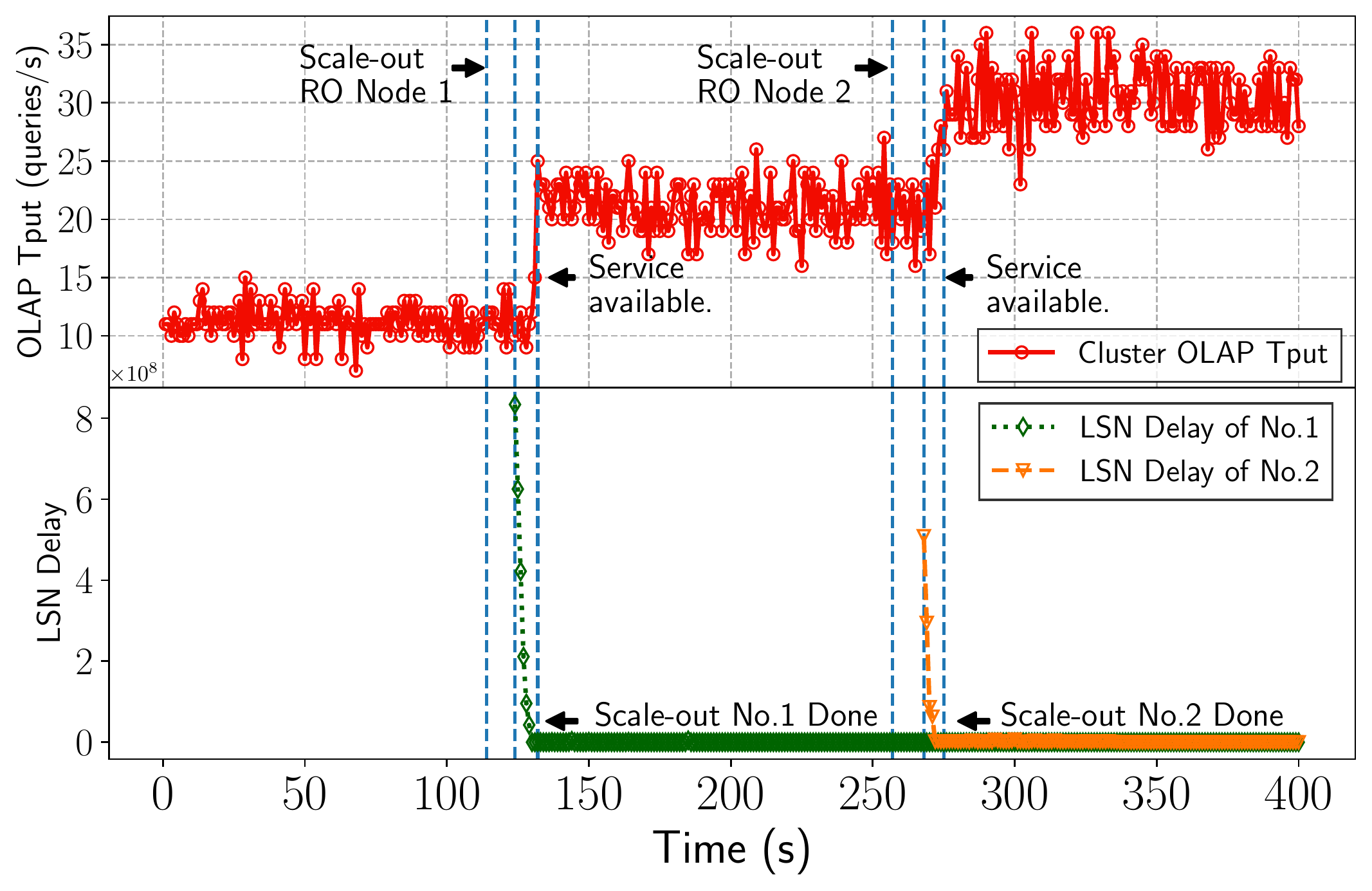}
		% \vspace{-4ex}
		% \vspace{1pt}
      \vspace{-2.5em}
\caption{Resource Elasticity on TPC-H}\label{fig:AP:scale}
     \vspace{-1em}
\end{figure}

\begin{table*}[t]
  \small
	\caption{Production workloads. The table describes different customer workload patterns.} \label{tab:realloads}
	\vspace{-1.0em}
	\setlength\tabcolsep{10pt} 
  \begin{tabular}{cccccccc}
    \hline
  \textbf{Workload}         & \textbf{DB Size}    & \textbf{Tables}   & \textbf{Max Table Size}   & \textbf{Avg.\# cols}   & \textbf{Queries}   & \textbf{Avg.\# joins}   & \textbf{ Avg.\# ops per plan} \\ \hline
  Cust1                     &  {2595.9 $GB$}           &  {997}             &  {393.3 $GB$}                &  {11.2}              &  {96}              &  {2.0}                &  {9.7}                   \\ 
  Cust2                     &  {163.2 $GB$}            &  {165}             &  {17.3 $GB$}                &  {27.2}             &  {311}              &  {1.3}                &  {10.0}                      \\  
  Cust3                     &  {736.2 $GB$}          &  {681}            &  {91.5 $GB$ }               &  {29.9}              &  {105}              &  {1.7}                &  {9.9}                  \\
  Cust4                     &  {47.8 $GB$}         &  {153}           &  {5.6 $GB$}                &  {13.5}              &  {106}              &  {9.0}               &  {41.9}                  \\ \hline
\end{tabular}
  % \vspace{-.5em}
%\vspace{-2em}
\end{table*}

\iffalse
\begin{figure*}[t]
	\centering
	\begin{subfigure}[t]{0.48\columnwidth}
		\centering
		\includegraphics[width=\columnwidth]{eval-figs/Speedup_Bucket_cust1.pdf}
		\vspace{-4ex}
		\subcaption{Cust1: Finance}\label{fig:speedup:cust1}
		% \vspace{1em}
    \end{subfigure}
	\begin{subfigure}[t]{0.48\columnwidth}
		\centering
		\includegraphics[width=\columnwidth]{eval-figs/Speedup_Bucket_cust2.pdf}
		\vspace{-4ex}
		\subcaption{Cust2: Logistics}\label{fig:speedup:cust2} 
    \end{subfigure}
	\begin{subfigure}[t]{0.48\columnwidth}
		\centering
		\includegraphics[width=\columnwidth]{eval-figs/Speedup_Bucket_cust3.pdf}
		\vspace{-4ex}
		\subcaption{Cust3: Video Marketing}\label{fig:speedup:cust3}
		% \vspace{1em}
    \end{subfigure}
	\begin{subfigure}[t]{0.48\columnwidth}
		\centering
		\includegraphics[width=\columnwidth]{eval-figs/Speedup_Bucket_cust4.pdf}
		\vspace{-4ex}
		% \vspace{1pt}
		\subcaption{Cust4: Gaming}\label{fig:speedup:cust4} 
    \end{subfigure}
     \vspace{-0.9em}
\caption{Distribution of speedups achieved by \sysname{} compared to row-based PolarDB on real-world applications.} \label{fig:speedup} 
     \vspace{-0.5em}
\end{figure*}
\fi

\begin{table}[h]
	\small
	\caption{\small  Distribution of queries at different IMCI speed-ups.}\label{tab:speedup}
	\vspace{-1.0em}

	\setlength\tabcolsep{12pt} 
	\begin{tabular}{c|cccc}
		\hline 
		\textbf{Speed-ups}  &   \textbf{Cust1}    &   \textbf{Cust2}  &   \textbf{Cust3}  &   \textbf{Cust4} \\ \hline
      [1, 2)     &   \textbf{55\%}     &   \textbf{67\%}     & 5\%      & 0\% \\ \hline
      [2, 5)     & 12\%     & 13\%     & 5\%      & 0\% \\ \hline
      [5, 10)    & 9\%      & 5\%      & 16\%     & 1\% \\ \hline
      [10, 100)  & 23\%     & 13\%     & 28\%     & 42\% \\ \hline
      [100, inf) & 1\%      & 2\%      &   \textbf{46\%}     &   \textbf{57\%} \\ \hline
\end{tabular}
	%\vspace{-25pt}
\end{table}

\begin{figure*}[t]
	\centering
	\begin{subfigure}[t]{0.48\columnwidth}
		\centering
		\includegraphics[width=\columnwidth]{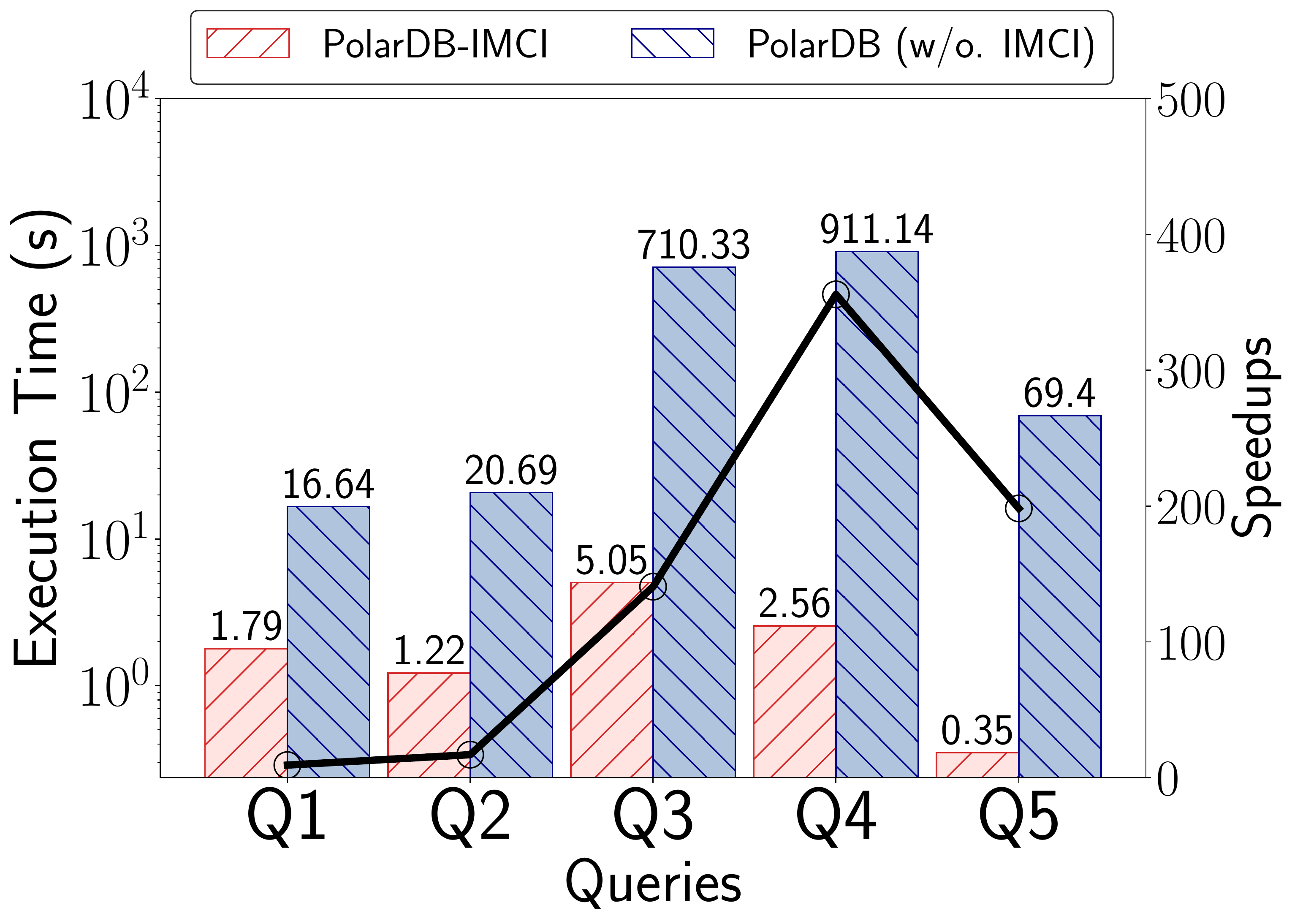}
		\vspace{-4ex}
		\subcaption{Cust1: Finance}
		% \vspace{1em}
    \end{subfigure}
	\begin{subfigure}[t]{0.48\columnwidth}
		\centering
		\includegraphics[width=\columnwidth]{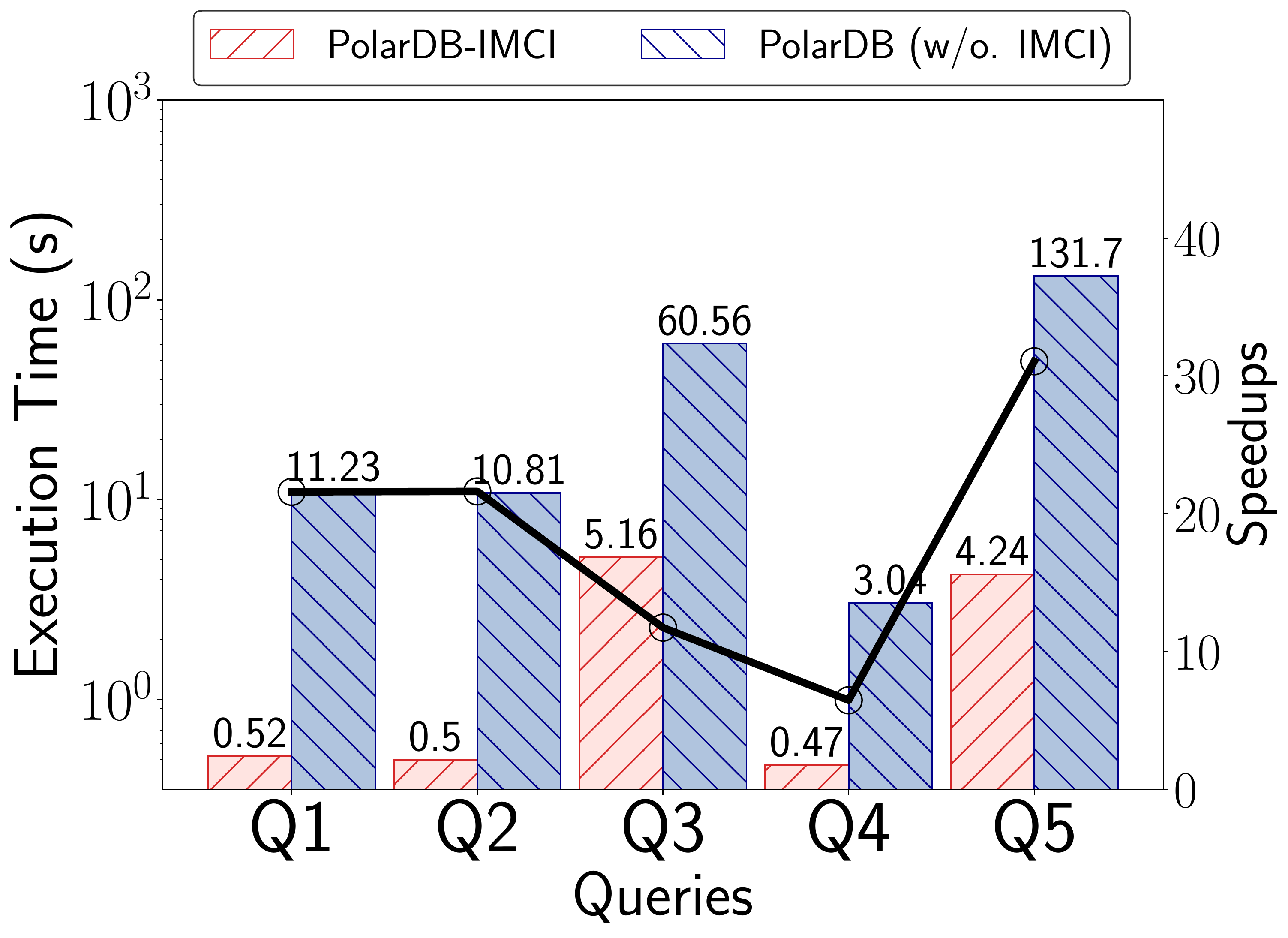}
		\vspace{-4ex}
		\subcaption{Cust2: Logistics}
    \end{subfigure}
	\begin{subfigure}[t]{0.48\columnwidth}
		\centering
		\includegraphics[width=\columnwidth]{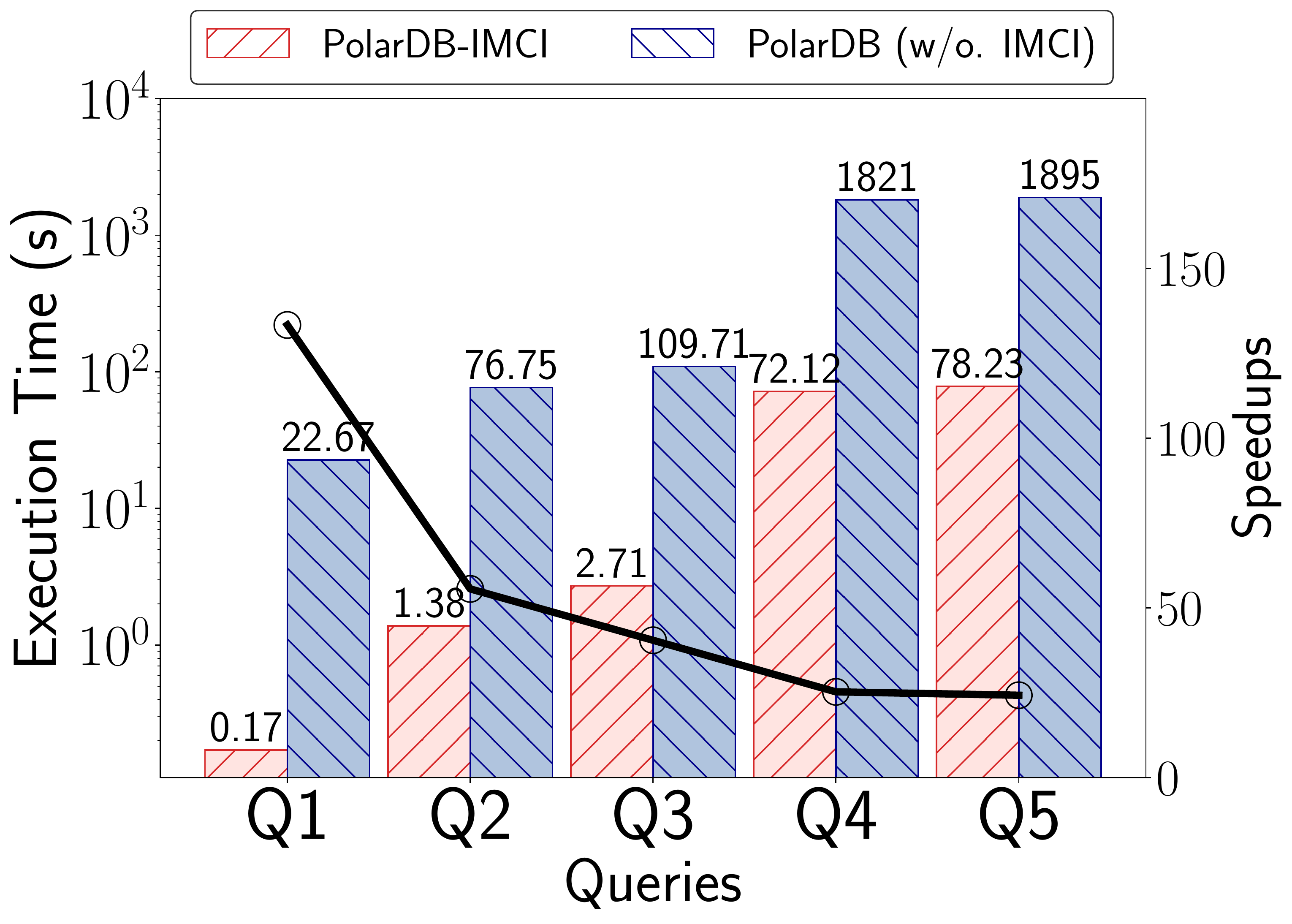}
		\vspace{-4ex}
		\subcaption{Cust3: Video Marketing}
		% \vspace{1em}
    \end{subfigure}
	\begin{subfigure}[t]{0.48\columnwidth}
		\centering
		\includegraphics[width=\columnwidth]{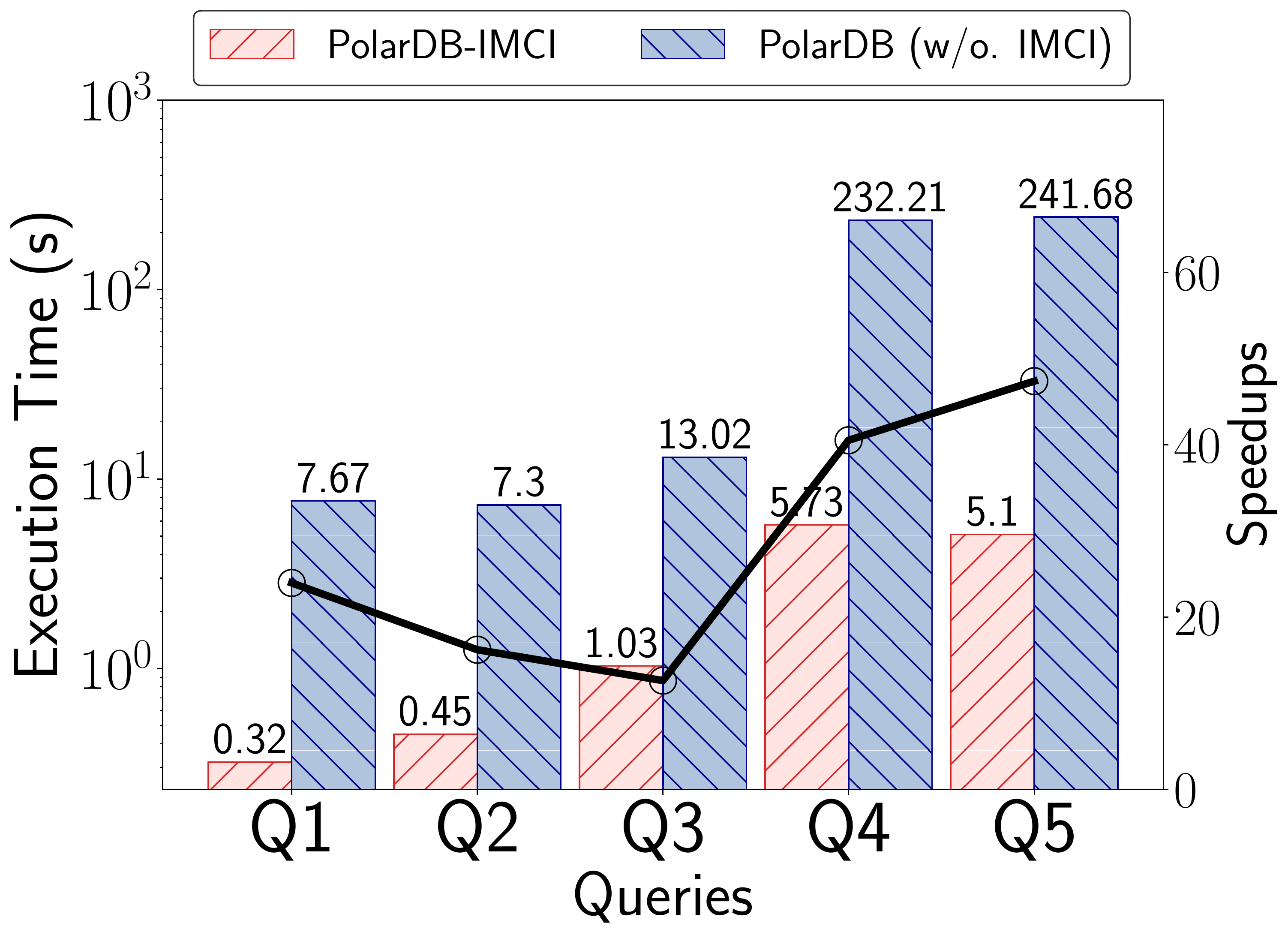}
		\vspace{-4ex}
		% \vspace{1pt}
		\subcaption{Cust4: Gaming}
    \end{subfigure}
     \vspace{-0.9em}
\caption{Speedups achieved by \sysname{} on representative queries. The left y-axis is in the log scale. }\label{fig:speedup2} 
     \vspace{-0.6em}
\end{figure*}

\begin{figure}[t]  
    \centering
    \includegraphics[width=1\columnwidth]{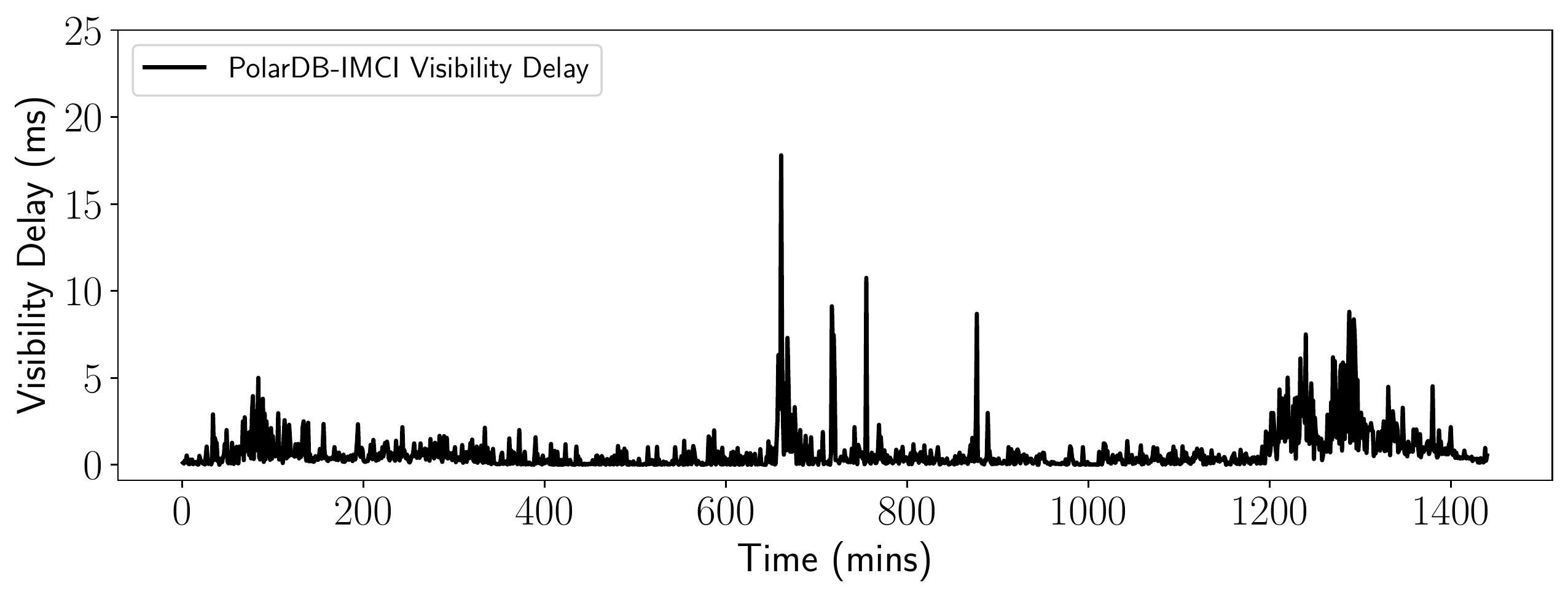}
    \vspace{-2.5em}
    \caption{\small Visibility Delay in real-world workloads.}\label{fig:realVD}
	\vspace{-1.5em}
\end{figure}

\subsection{Data Freshness}\label{sec:eval:freshness}
Data freshness (i.e., \textbf{G\#4}) is critical to the quality of analytical results. 
We evaluated data freshness by visibility delay (VD), which is the time taken for an update committed on an RW node to be readable on RO nodes~\cite{tidb2020, chen2022bytehtap}.
\fig{fig:freshness} provides the results of VD at different percentiles on TPC-C workloads with data warehouses = 100. 
% Overall, write operations incurred longer VD compared to insert operations.
% , and the maximum of VD was less than 160$ms$. 
% The reason is straightforward, \sysname{} implemented out-place updates (\chref{sec:column}), i.e., a write was performed by a delete and an insert.
%  \sysname{} achieved at most $\sim$$126k$ tps and $\sim$$45k$ tps 
% for sysbench insert-only and write-only workloads respectively by using 1024 OLTP client threads. 
\sysname{} achieved low visibility delay for three reasons: first, CALS (\chref{subsec:CALS}) minimized the update propagation window;
second, the updates on column indexes are out-place and lightweight (\chref{sec:column:layout}); third, RDMA-equipped PolarFS reduced the shipping time.
% As we analyze different latencies percentiles, VD increases more significant at higher percentiles. This is because 

% Furthermore, to understand the VD in detail, we collected the breakdown data of VD for each workload at 99.9th. 
% % \fig{fig:VD:insert:tput} and \fig{fig:VD:write:tput} show the results. 
% For both two kinds of workloads, the time for log shipping 
% contributed to the major part of VD ($98.39$\%$\sim$$99.86\%$). It means the components of IMCI did not cause much overhead on VD, and
% VD was bounded by the cross-node shipping latency. 
% Log shipping latency increased with the number of clients because the volume of REDO logs increased and required much more network packages and I/O.
% However, thanks to our advanced infrastructures on the cloud (e.g., RDMA), the cost for in-database log shipping was significantly lower than other on-premise databases, 
% and outperformed to message-queue-based approaches (e.g., Flink~\cite{flink}). 

\noindent\textbf{Effectiveness of parallel replay.} 
To provide additional support for the claim that components of column indexes should never be the bottleneck, 
we tested each component of \sysname{} individually, and report the maximum throughput on each component with varying threads. 
During the experiment, we used 512 OLTP clients to saturate OLTP throughput on the TPC-C workload and achieved $1934.97$$tps$ (i.e., $116098$$tpmC$) throughput.
%and achieved $\sim$$34$$k$ tps throughput at the maximum  
%on sysbench write-only workload. 
\fig{fig:propagation:AP} shows the results. 
The maximum throughput of updating the RID locator and data Packs is much higher 
than the maximum throughput of OLTP on the RW node ($\times30.2$ to $\times61.3$). 
Besides, replaying REDO logs on a row-based buffer pool is not the bottleneck.
We also test the maximum throughput of physical log parsing (per thread) and committing. 
The throughputs are $\sim$$34k$ and $\sim$$459k$ respectively, which is also significantly higher. 
% Arguement: Replay Throughput of \sysname is adequate. \fig{fig:propagation:AP}

\subsection{Resource Elasticity}\label{sec:eval:elasticity}
The desiderata on resource elasticity (i.e., \textbf{G\#5}) drives our cloud-native implementation.  
To test the elasticity of \sysname{}, we used sysbench insert-only workloads with 3900 insertions per second for the TP workload
and TPC-H for the AP workload.
%\fig{fig:AP:scaleup} shows the results on OLAP throughput when the given CPU cores increased. 
%Since Alibaba Cloud provides virtual CPUs over physical cores, \sysname{} can be upgraded automatically in sub-seconds. 

To scale out (i.e., add new RO nodes), \sysname{} relies on the checkpoints technique (\chref{sec:fast_startup}) for a fast start-up. We used TPC-H Q6 for scaling our experiments.
%In our evaluation, we used the checkpoints with the size of $\sim$$30$$GB$ to test the elasticity time. \fig{fig:AP:scale} shows the results. 
We added the first new RO node into the cluster at 114s. It took 10s for \sysname{} to build in-memory components from the checkpoints. 
At 124s, when the newly added RO node (i.e., No.1) was able to serve the new incoming OLAP requests, the proxy server balanced the 
traffic and started new sessions to No.1. Thus, the cluster's OLAP throughput increased incrementally (see the top part of \fig{fig:AP:scale}). However, at the beginning of the start-up, 
the LSN delay of No.1 was extremely high (see the bottom part of \fig{fig:AP:scale}) since the new node still needed to catch up on updates committed after the checkpoint.
Thanks to our high-performance updates propagation framework, No.1 could catch up to the latest state in a short time (9s). 
At 133s, No.1 could behave as a normal RO node to serve OLAP requests. 
We then added another new RO node (i.e., No.2) to the cluster at 257s. 
No.2 was able to provide services at 268s and could catch up to the latest at 276s. 
Notably, No.2 took less time to catch up to RW than No.1 since No.2 started from the next round checkpoint.

Overall, \sysname{} achieved strong elasticity: it takes tens of seconds to scale out.

\subsection{Performance of Production Deployment}\label{sec:eval:production}
In the last experiment, we studied \sysname{}'s performance on several real-world customer
workloads in a production environment. These workloads represent four diverse real-time applications where HTAP is highly desirable, 
i.e., finance, logistics, video marketing, and online gaming. \tab{tab:realloads} reports some aggregate statistics
about the schema of these workloads. Generally, these customer workloads represent complex query patterns over diverse data schemas and database sizes. 
\tab{tab:speedup} shows the distribution of speed-ups achieved by \sysname{} compared to row-based PolarDB and \fig{fig:speedup2} shows the representative queries. It revealed that column indexes can result in orders of magnitude performance gains for slow SQL queries.
% \fig{fig:speedup2} shows the execution time of the top-five slow SQL that was executed on row-based PolarDB, and their corresponding execution time on \sysname. 
% Generally, these customer queries represent complex queries over diverse data schemas and database sizes.
% We also calculated the speed-ups, which were shown in the right y-axis.
% In summary, it revealed that column indexes can result in orders of magnitude performance gains for slow SQL queries. 

We then monitored the visibility delay between RW and RO nodes. The results are shown in \fig{fig:realVD}. During 24 hours, the visibility delay was changed with the customer's OLTP throughput and was always \textless$20ms$.
\section{Conclusion}\label{sec:conclusion}
This paper present \sysname{}, a cloud-native HTAP database that achieves advanced 
OLAP performance with minimal perturbation on OLTP, and optimized visibility delay for better data freshness.
Our evaluation results show that \sysname{} can handle 
hybrid workloads efficiently in both experimental and productional environments.

%\section{ACKNOWLEDGMENTS}\label{sec:acknowledgments}  
\begin{acks}
 \sysname{} owes a great deal to our customers, whose feedback and suggestions were instrumental in the design of its architecture. 
 We gratefully acknowledge the contributions of Ming Zhao, XuDong Wu, HuaWei Xue, and Shuai Jiang to the development of \sysname{}. 
 Additionally, we extend heartfelt gratitude to the anonymous reviewers whose valuable comments greatly improved this paper.
\end{acks}

\bibliographystyle{ACM-Reference-Format}
\bibliography{ref}


\begin{thebibliography}{56}

%%% the Web address.

\ifx \showCODEN    \undefined \def \showCODEN     #1{\unskip}     \fi
\ifx \showDOI      \undefined \def \showDOI       #1{#1}\fi
\ifx \showISBNx    \undefined \def \showISBNx     #1{\unskip}     \fi
\ifx \showISBNxiii \undefined \def \showISBNxiii  #1{\unskip}     \fi
\ifx \showISSN     \undefined \def \showISSN      #1{\unskip}     \fi
\ifx \showLCCN     \undefined \def \showLCCN      #1{\unskip}     \fi
\ifx \shownote     \undefined \def \shownote      #1{#1}          \fi
\ifx \showarticletitle \undefined \def \showarticletitle #1{#1}   \fi
\ifx \showURL      \undefined \def \showURL       {\relax}        \fi
% invisible to TeX
\providecommand\bibfield[2]{#2}
\providecommand\bibinfo[2]{#2}
\providecommand\natexlab[1]{#1}
\providecommand\showeprint[2][]{arXiv:#2}

\bibitem[\protect\citeauthoryear{Balkesen, Teubner, Alonso, and
  {\"{O}}zsu}{Balkesen et~al\mbox{.}}{2013}]%
        {BalkesenTAO13ParJoin}
\bibfield{author}{\bibinfo{person}{Cagri Balkesen}, \bibinfo{person}{Jens
  Teubner}, \bibinfo{person}{Gustavo Alonso}, {and} \bibinfo{person}{M.~Tamer
  {\"{O}}zsu}.} \bibinfo{year}{2013}\natexlab{}.
\newblock \showarticletitle{Main-memory hash joins on multi-core CPUs: Tuning
  to the underlying hardware}. In \bibinfo{booktitle}{\emph{29th {IEEE}
  International Conference on Data Engineering, {ICDE} 2013, Brisbane,
  Australia, April 8-12, 2013}}. \bibinfo{publisher}{{IEEE} Computer Society},
  \bibinfo{pages}{362--373}.
\newblock


\bibitem[\protect\citeauthoryear{Barber, Huras, Lohman, Mohan, Mueller,
  {\"O}zcan, Pirahesh, Raman, Sidle, Sidorkin, et~al\mbox{.}}{Barber
  et~al\mbox{.}}{2016}]%
        {barber2016wildfire}
\bibfield{author}{\bibinfo{person}{Ronald Barber}, \bibinfo{person}{Matt
  Huras}, \bibinfo{person}{Guy Lohman}, \bibinfo{person}{C Mohan},
  \bibinfo{person}{Rene Mueller}, \bibinfo{person}{Fatma {\"O}zcan},
  \bibinfo{person}{Hamid Pirahesh}, \bibinfo{person}{Vijayshankar Raman},
  \bibinfo{person}{Richard Sidle}, \bibinfo{person}{Oleg Sidorkin},
  {et~al\mbox{.}}} \bibinfo{year}{2016}\natexlab{}.
\newblock \showarticletitle{Wildfire: Concurrent blazing data ingest and
  analytics}. In \bibinfo{booktitle}{\emph{Proceedings of the 2016
  International Conference on Management of Data}}.
  \bibinfo{pages}{2077--2080}.
\newblock


\bibitem[\protect\citeauthoryear{Barber, Lohman, Pandis, Raman, Sidle,
  Attaluri, Chainani, Lightstone, and Sharpe}{Barber et~al\mbox{.}}{2014}]%
        {BarberLPRSACLS14CHT}
\bibfield{author}{\bibinfo{person}{Ronald Barber}, \bibinfo{person}{Guy~M.
  Lohman}, \bibinfo{person}{Ippokratis Pandis}, \bibinfo{person}{Vijayshankar
  Raman}, \bibinfo{person}{Richard Sidle}, \bibinfo{person}{Gopi~K. Attaluri},
  \bibinfo{person}{Naresh Chainani}, \bibinfo{person}{Sam Lightstone}, {and}
  \bibinfo{person}{David Sharpe}.} \bibinfo{year}{2014}\natexlab{}.
\newblock \showarticletitle{Memory-Efficient Hash Joins}.
\newblock \bibinfo{journal}{\emph{Proc. {VLDB} Endow.}} \bibinfo{volume}{8},
  \bibinfo{number}{4} (\bibinfo{year}{2014}), \bibinfo{pages}{353--364}.
\newblock


\bibitem[\protect\citeauthoryear{Boncz, Zukowski, and Nes}{Boncz
  et~al\mbox{.}}{2005}]%
        {BonczZN05MonetDB}
\bibfield{author}{\bibinfo{person}{Peter~A. Boncz}, \bibinfo{person}{Marcin
  Zukowski}, {and} \bibinfo{person}{Niels Nes}.}
  \bibinfo{year}{2005}\natexlab{}.
\newblock \showarticletitle{MonetDB/X100: Hyper-Pipelining Query Execution}. In
  \bibinfo{booktitle}{\emph{Second Biennial Conference on Innovative Data
  Systems Research, {CIDR} 2005, Asilomar, CA, USA, January 4-7, 2005, Online
  Proceedings}}. \bibinfo{publisher}{www.cidrdb.org},
  \bibinfo{pages}{225--237}.
\newblock


\bibitem[\protect\citeauthoryear{Borthakur, Gray, Sarma, Muthukkaruppan,
  Spiegelberg, Kuang, Ranganathan, Molkov, Menon, Rash, Schmidt, and
  Aiyer}{Borthakur et~al\mbox{.}}{2011}]%
        {fackbook}
\bibfield{author}{\bibinfo{person}{Dhruba Borthakur}, \bibinfo{person}{Jonathan
  Gray}, \bibinfo{person}{Joydeep~Sen Sarma}, \bibinfo{person}{Kannan
  Muthukkaruppan}, \bibinfo{person}{Nicolas Spiegelberg},
  \bibinfo{person}{Hairong Kuang}, \bibinfo{person}{Karthik Ranganathan},
  \bibinfo{person}{Dmytro Molkov}, \bibinfo{person}{Aravind Menon},
  \bibinfo{person}{Samuel Rash}, \bibinfo{person}{Rodrigo Schmidt}, {and}
  \bibinfo{person}{Amitanand Aiyer}.} \bibinfo{year}{2011}\natexlab{}.
\newblock \showarticletitle{Apache Hadoop Goes Realtime at Facebook}. In
  \bibinfo{booktitle}{\emph{Proceedings of the 2011 ACM SIGMOD International
  Conference on Management of Data}} \emph{(\bibinfo{series}{SIGMOD '11})}.
  \bibinfo{publisher}{Association for Computing Machinery},
  \bibinfo{address}{New York, NY, USA}, \bibinfo{pages}{1071–1080}.
\newblock
\showISBNx{9781450306614}


\bibitem[\protect\citeauthoryear{Butterstein, Martin, Stolze, Beier, Zhong, and
  Wang}{Butterstein et~al\mbox{.}}{2020}]%
        {butterstein2020replication}
\bibfield{author}{\bibinfo{person}{Dennis Butterstein}, \bibinfo{person}{Daniel
  Martin}, \bibinfo{person}{Knut Stolze}, \bibinfo{person}{Felix Beier},
  \bibinfo{person}{Jia Zhong}, {and} \bibinfo{person}{Lingyun Wang}.}
  \bibinfo{year}{2020}\natexlab{}.
\newblock \showarticletitle{Replication at the speed of change: a fast,
  scalable replication solution for near real-time HTAP processing}.
\newblock \bibinfo{journal}{\emph{Proceedings of the VLDB Endowment}}
  \bibinfo{volume}{13}, \bibinfo{number}{12} (\bibinfo{year}{2020}),
  \bibinfo{pages}{3245--3257}.
\newblock


\bibitem[\protect\citeauthoryear{Cao, Yang, Chen, Zhou, Li, and Qi}{Cao
  et~al\mbox{.}}{2019}]%
        {frauddete}
\bibfield{author}{\bibinfo{person}{Shaosheng Cao}, \bibinfo{person}{XinXing
  Yang}, \bibinfo{person}{Cen Chen}, \bibinfo{person}{Jun Zhou},
  \bibinfo{person}{Xiaolong Li}, {and} \bibinfo{person}{Yuan Qi}.}
  \bibinfo{year}{2019}\natexlab{}.
\newblock \showarticletitle{TitAnt: Online Real-Time Transaction Fraud
  Detection in Ant Financial}.
\newblock \bibinfo{journal}{\emph{Proc. VLDB Endow.}} \bibinfo{volume}{12},
  \bibinfo{number}{12} (\bibinfo{date}{aug} \bibinfo{year}{2019}),
  \bibinfo{pages}{2082–2093}.
\newblock
\showISSN{2150-8097}


\bibitem[\protect\citeauthoryear{Cao, Liu, Wang, Chen, Zhu, Zheng, Wang, and
  Ma}{Cao et~al\mbox{.}}{2018}]%
        {Cao2018}
\bibfield{author}{\bibinfo{person}{Wei Cao}, \bibinfo{person}{Zhenjun Liu},
  \bibinfo{person}{Peng Wang}, \bibinfo{person}{Sen Chen},
  \bibinfo{person}{Caifeng Zhu}, \bibinfo{person}{Song Zheng},
  \bibinfo{person}{Yuhui Wang}, {and} \bibinfo{person}{Guoqing Ma}.}
  \bibinfo{year}{2018}\natexlab{}.
\newblock \showarticletitle{{PolarFS: An ultralow latency and failure resilient
  distributed file system for shared storage cloud database}}.
\newblock \bibinfo{journal}{\emph{Proceedings of the VLDB Endowment}}
  \bibinfo{volume}{11}, \bibinfo{number}{12} (\bibinfo{year}{2018}),
  \bibinfo{pages}{1849--1862}.
\newblock
\showISSN{21508097}


\bibitem[\protect\citeauthoryear{Cao, Zhang, Yang, Li, Wang, Hu, Cheng, Chen,
  Liu, Fang, Wang, Wang, Sun, Yang, Cheng, Chen, Wu, Hu, Zhao, Gao, Cai, Zhang,
  and Tong}{Cao et~al\mbox{.}}{2021}]%
        {cao2021polardb}
\bibfield{author}{\bibinfo{person}{Wei Cao}, \bibinfo{person}{Yingqiang Zhang},
  \bibinfo{person}{Xinjun Yang}, \bibinfo{person}{Feifei Li},
  \bibinfo{person}{Sheng Wang}, \bibinfo{person}{Qingda Hu},
  \bibinfo{person}{Xuntao Cheng}, \bibinfo{person}{Zongzhi Chen},
  \bibinfo{person}{Zhenjun Liu}, \bibinfo{person}{Jing Fang},
  \bibinfo{person}{Bo Wang}, \bibinfo{person}{Yuhui Wang},
  \bibinfo{person}{Haiqing Sun}, \bibinfo{person}{Ze Yang},
  \bibinfo{person}{Zhushi Cheng}, \bibinfo{person}{Sen Chen},
  \bibinfo{person}{Jian Wu}, \bibinfo{person}{Wei Hu}, \bibinfo{person}{Jianwei
  Zhao}, \bibinfo{person}{Yusong Gao}, \bibinfo{person}{Songlu Cai},
  \bibinfo{person}{Yunyang Zhang}, {and} \bibinfo{person}{Jiawang Tong}.}
  \bibinfo{year}{2021}\natexlab{}.
\newblock \showarticletitle{PolarDB Serverless: {A} Cloud Native Database for
  Disaggregated Data Centers}. In \bibinfo{booktitle}{\emph{{SIGMOD} '21:
  International Conference on Management of Data, Virtual Event, China, June
  20-25, 2021}}. \bibinfo{publisher}{{ACM}}, \bibinfo{pages}{2477--2489}.
\newblock


\bibitem[\protect\citeauthoryear{Chaudhuri, Das, and Srivastava}{Chaudhuri
  et~al\mbox{.}}{2004}]%
        {ChaudhuriDS04}
\bibfield{author}{\bibinfo{person}{Surajit Chaudhuri}, \bibinfo{person}{Gautam
  Das}, {and} \bibinfo{person}{Utkarsh Srivastava}.}
  \bibinfo{year}{2004}\natexlab{}.
\newblock \showarticletitle{Effective Use of Block-Level Sampling in Statistics
  Estimation}. In \bibinfo{booktitle}{\emph{Proceedings of the {ACM} {SIGMOD}
  International Conference on Management of Data, Paris, France, June 13-18,
  2004}}. \bibinfo{publisher}{{ACM}}, \bibinfo{pages}{287--298}.
\newblock


\bibitem[\protect\citeauthoryear{Chaudhuri, Motwani, and Narasayya}{Chaudhuri
  et~al\mbox{.}}{1998}]%
        {ChaudhuriMN98}
\bibfield{author}{\bibinfo{person}{Surajit Chaudhuri}, \bibinfo{person}{Rajeev
  Motwani}, {and} \bibinfo{person}{Vivek~R. Narasayya}.}
  \bibinfo{year}{1998}\natexlab{}.
\newblock \showarticletitle{Random Sampling for Histogram Construction: How
  much is enough?}. In \bibinfo{booktitle}{\emph{{SIGMOD} 1998, Proceedings
  {ACM} {SIGMOD} International Conference on Management of Data, June 2-4,
  1998, Seattle, Washington, {USA}}}. \bibinfo{publisher}{{ACM} Press},
  \bibinfo{pages}{436--447}.
\newblock


\bibitem[\protect\citeauthoryear{Chen, Ding, Liu, Li, Zhang, Zhang, Wei, Cao,
  Zou, Liu, et~al\mbox{.}}{Chen et~al\mbox{.}}{2022a}]%
        {chen2022bytehtap}
\bibfield{author}{\bibinfo{person}{Jianjun Chen}, \bibinfo{person}{Yonghua
  Ding}, \bibinfo{person}{Ye Liu}, \bibinfo{person}{Fangshi Li},
  \bibinfo{person}{Li Zhang}, \bibinfo{person}{Mingyi Zhang},
  \bibinfo{person}{Kui Wei}, \bibinfo{person}{Lixun Cao}, \bibinfo{person}{Dan
  Zou}, \bibinfo{person}{Yang Liu}, {et~al\mbox{.}}}
  \bibinfo{year}{2022}\natexlab{a}.
\newblock \showarticletitle{ByteHTAP: bytedance's HTAP system with high data
  freshness and strong data consistency}.
\newblock \bibinfo{journal}{\emph{Proceedings of the VLDB Endowment}}
  \bibinfo{volume}{15}, \bibinfo{number}{12} (\bibinfo{year}{2022}),
  \bibinfo{pages}{3411--3424}.
\newblock


\bibitem[\protect\citeauthoryear{Chen, Ailamaki, Gibbons, and Mowry}{Chen
  et~al\mbox{.}}{2004}]%
        {ChenAGM04Prefetch}
\bibfield{author}{\bibinfo{person}{Shimin Chen}, \bibinfo{person}{Anastassia
  Ailamaki}, \bibinfo{person}{Phillip~B. Gibbons}, {and}
  \bibinfo{person}{Todd~C. Mowry}.} \bibinfo{year}{2004}\natexlab{}.
\newblock \showarticletitle{Improving Hash Join Performance through
  Prefetching}. In \bibinfo{booktitle}{\emph{Proceedings of the 20th
  International Conference on Data Engineering, {ICDE} 2004, 30 March - 2 April
  2004, Boston, MA, {USA}}}. \bibinfo{publisher}{{IEEE} Computer Society},
  \bibinfo{pages}{116--127}.
\newblock


\bibitem[\protect\citeauthoryear{Chen, Yang, Li, Cheng, Hu, Miao, Xie, Wu,
  Wang, Song, et~al\mbox{.}}{Chen et~al\mbox{.}}{2022b}]%
        {chen2022cloudjump}
\bibfield{author}{\bibinfo{person}{Zongzhi Chen}, \bibinfo{person}{Xinjun
  Yang}, \bibinfo{person}{Feifei Li}, \bibinfo{person}{Xuntao Cheng},
  \bibinfo{person}{Qingda Hu}, \bibinfo{person}{Zheyu Miao},
  \bibinfo{person}{Rongbiao Xie}, \bibinfo{person}{Xiaofei Wu},
  \bibinfo{person}{Kang Wang}, \bibinfo{person}{Zhao Song}, {et~al\mbox{.}}}
  \bibinfo{year}{2022}\natexlab{b}.
\newblock \showarticletitle{CloudJump: optimizing cloud databases for cloud
  storages}.
\newblock \bibinfo{journal}{\emph{Proceedings of the VLDB Endowment}}
  \bibinfo{volume}{15}, \bibinfo{number}{12} (\bibinfo{year}{2022}),
  \bibinfo{pages}{3432--3444}.
\newblock


\bibitem[\protect\citeauthoryear{ClickHouse}{ClickHouse}{[n.d.]}]%
        {ckjoin}
\bibfield{author}{\bibinfo{person}{Inc. ClickHouse}.}
  \bibinfo{year}{[n.d.]}\natexlab{}.
\newblock \bibinfo{title}{{ClickHouse — Roadmap 2023.}}
\newblock
  \bibinfo{howpublished}{\url{https://github.com/ClickHouse/ClickHouse/issues/44767}}.
\newblock


\bibitem[\protect\citeauthoryear{ClickHouse}{ClickHouse}{2022}]%
        {ck}
\bibfield{author}{\bibinfo{person}{Inc. ClickHouse}.}
  \bibinfo{year}{2022}\natexlab{}.
\newblock \bibinfo{title}{{ClickHouse — open source distributed
  column-oriented DBMS.}}
\newblock
  \bibinfo{howpublished}{\url{https://github.com/ClickHouse/ClickHouse/tree/22.6}}.
\newblock


\bibitem[\protect\citeauthoryear{Cole, Funke, Giakoumakis, Guy, Kemper,
  Krompass, Kuno, Nambiar, Neumann, Poess, Sattler, Seibold, Simon, and
  Waas}{Cole et~al\mbox{.}}{2011}]%
        {cole2011mixed}
\bibfield{author}{\bibinfo{person}{Richard~L. Cole}, \bibinfo{person}{Florian
  Funke}, \bibinfo{person}{Leo Giakoumakis}, \bibinfo{person}{Wey Guy},
  \bibinfo{person}{Alfons Kemper}, \bibinfo{person}{Stefan Krompass},
  \bibinfo{person}{Harumi~A. Kuno}, \bibinfo{person}{Raghunath~Othayoth
  Nambiar}, \bibinfo{person}{Thomas Neumann}, \bibinfo{person}{Meikel Poess},
  \bibinfo{person}{Kai{-}Uwe Sattler}, \bibinfo{person}{Michael Seibold},
  \bibinfo{person}{Eric Simon}, {and} \bibinfo{person}{Florian Waas}.}
  \bibinfo{year}{2011}\natexlab{}.
\newblock \showarticletitle{The mixed workload CH-benCHmark}. In
  \bibinfo{booktitle}{\emph{Proceedings of the Fourth International Workshop on
  Testing Database Systems, DBTest 2011, Athens, Greece, June 13, 2011}}.
  \bibinfo{publisher}{{ACM}}, \bibinfo{pages}{8}.
\newblock


\bibitem[\protect\citeauthoryear{Community.}{Community.}{2023}]%
        {flink}
\bibfield{author}{\bibinfo{person}{Apache Community.}}
  \bibinfo{year}{2023}\natexlab{}.
\newblock \bibinfo{title}{{Apache Flink.}}
\newblock \bibinfo{howpublished}{\url{https://flink.apache.org/}}.
\newblock


\bibitem[\protect\citeauthoryear{COUNCIL}{COUNCIL}{2014}]%
        {tpcc}
\bibfield{author}{\bibinfo{person}{THE TRANSACTION~PROCESSING COUNCIL}.}
  \bibinfo{year}{2014}\natexlab{}.
\newblock \bibinfo{title}{{TPC-C}}.
\newblock \bibinfo{howpublished}{\url{http://www.tpc.org/tpcc/}}.
\newblock


\bibitem[\protect\citeauthoryear{Dageville, Cruanes, Zukowski, Antonov, Avanes,
  Bock, Claybaugh, Engovatov, Hentschel, Huang, et~al\mbox{.}}{Dageville
  et~al\mbox{.}}{2016}]%
        {dageville2016snowflake}
\bibfield{author}{\bibinfo{person}{Benoit Dageville}, \bibinfo{person}{Thierry
  Cruanes}, \bibinfo{person}{Marcin Zukowski}, \bibinfo{person}{Vadim Antonov},
  \bibinfo{person}{Artin Avanes}, \bibinfo{person}{Jon Bock},
  \bibinfo{person}{Jonathan Claybaugh}, \bibinfo{person}{Daniel Engovatov},
  \bibinfo{person}{Martin Hentschel}, \bibinfo{person}{Jiansheng Huang},
  {et~al\mbox{.}}} \bibinfo{year}{2016}\natexlab{}.
\newblock \showarticletitle{The snowflake elastic data warehouse}. In
  \bibinfo{booktitle}{\emph{Proceedings of the 2016 International Conference on
  Management of Data}}. \bibinfo{pages}{215--226}.
\newblock


\bibitem[\protect\citeauthoryear{Deng, Gao, and Vuppalapati}{Deng
  et~al\mbox{.}}{2015}]%
        {marketing}
\bibfield{author}{\bibinfo{person}{Lei Deng}, \bibinfo{person}{Jerry Gao},
  {and} \bibinfo{person}{Chandrasekar Vuppalapati}.}
  \bibinfo{year}{2015}\natexlab{}.
\newblock \showarticletitle{Building a Big Data Analytics Service Framework for
  Mobile Advertising and Marketing}. In \bibinfo{booktitle}{\emph{First {IEEE}
  International Conference on Big Data Computing Service and Applications,
  BigDataService 2015, Redwood City, CA, USA, March 30 - April 2, 2015}}.
  \bibinfo{publisher}{{IEEE} Computer Society}, \bibinfo{pages}{256--266}.
\newblock


\bibitem[\protect\citeauthoryear{Depoutovitch, Chen, Chen, Larson, Lin, Ng,
  Cui, Liu, Huang, Xiao, et~al\mbox{.}}{Depoutovitch et~al\mbox{.}}{2020}]%
        {depoutovitch2020taurus}
\bibfield{author}{\bibinfo{person}{Alex Depoutovitch}, \bibinfo{person}{Chong
  Chen}, \bibinfo{person}{Jin Chen}, \bibinfo{person}{Paul Larson},
  \bibinfo{person}{Shu Lin}, \bibinfo{person}{Jack Ng}, \bibinfo{person}{Wenlin
  Cui}, \bibinfo{person}{Qiang Liu}, \bibinfo{person}{Wei Huang},
  \bibinfo{person}{Yong Xiao}, {et~al\mbox{.}}}
  \bibinfo{year}{2020}\natexlab{}.
\newblock \showarticletitle{Taurus database: How to be fast, available, and
  frugal in the cloud}. In \bibinfo{booktitle}{\emph{Proceedings of the 2020
  ACM SIGMOD International Conference on Management of Data}}.
  \bibinfo{pages}{1463--1478}.
\newblock


\bibitem[\protect\citeauthoryear{Driscoll, Sarnak, Sleator, and
  Tarjan}{Driscoll et~al\mbox{.}}{1989}]%
        {driscoll1989making}
\bibfield{author}{\bibinfo{person}{James~R Driscoll}, \bibinfo{person}{Neil
  Sarnak}, \bibinfo{person}{Daniel~D Sleator}, {and} \bibinfo{person}{Robert~E
  Tarjan}.} \bibinfo{year}{1989}\natexlab{}.
\newblock \showarticletitle{Making data structures persistent}.
\newblock \bibinfo{journal}{\emph{Journal of computer and system sciences}}
  \bibinfo{volume}{38}, \bibinfo{number}{1} (\bibinfo{year}{1989}),
  \bibinfo{pages}{86--124}.
\newblock


\bibitem[\protect\citeauthoryear{F{\"{a}}rber, May, Lehner, Gro{\ss}e,
  M{\"{u}}ller, Rauhe, and Dees}{F{\"{a}}rber et~al\mbox{.}}{2012}]%
        {FarberMLGMRD12HANA}
\bibfield{author}{\bibinfo{person}{Franz F{\"{a}}rber}, \bibinfo{person}{Norman
  May}, \bibinfo{person}{Wolfgang Lehner}, \bibinfo{person}{Philipp Gro{\ss}e},
  \bibinfo{person}{Ingo M{\"{u}}ller}, \bibinfo{person}{Hannes Rauhe}, {and}
  \bibinfo{person}{Jonathan Dees}.} \bibinfo{year}{2012}\natexlab{}.
\newblock \showarticletitle{The {SAP} {HANA} Database -- An Architecture
  Overview}.
\newblock \bibinfo{journal}{\emph{{IEEE} Data Eng. Bull.}}
  \bibinfo{volume}{35}, \bibinfo{number}{1} (\bibinfo{year}{2012}),
  \bibinfo{pages}{28--33}.
\newblock


\bibitem[\protect\citeauthoryear{Gupta, Agarwal, Tan, Kulesza, Pathak, Stefani,
  and Srinivasan}{Gupta et~al\mbox{.}}{2015}]%
        {gupta2015amazon}
\bibfield{author}{\bibinfo{person}{Anurag Gupta}, \bibinfo{person}{Deepak
  Agarwal}, \bibinfo{person}{Derek Tan}, \bibinfo{person}{Jakub Kulesza},
  \bibinfo{person}{Rahul Pathak}, \bibinfo{person}{Stefano Stefani}, {and}
  \bibinfo{person}{Vidhya Srinivasan}.} \bibinfo{year}{2015}\natexlab{}.
\newblock \showarticletitle{Amazon redshift and the case for simpler data
  warehouses}. In \bibinfo{booktitle}{\emph{Proceedings of the 2015 ACM SIGMOD
  international conference on management of data}}.
  \bibinfo{pages}{1917--1923}.
\newblock


\bibitem[\protect\citeauthoryear{Haas and Stokes}{Haas and Stokes}{1998}]%
        {haas1998estimating}
\bibfield{author}{\bibinfo{person}{Peter~J Haas} {and} \bibinfo{person}{Lynne
  Stokes}.} \bibinfo{year}{1998}\natexlab{}.
\newblock \showarticletitle{Estimating the number of classes in a finite
  population}.
\newblock \bibinfo{journal}{\emph{J. Amer. Statist. Assoc.}}
  \bibinfo{volume}{93}, \bibinfo{number}{444} (\bibinfo{year}{1998}),
  \bibinfo{pages}{1475--1487}.
\newblock


\bibitem[\protect\citeauthoryear{Huang, Liu, Cui, Fang, Ma, Xu, Shen, Tang,
  Zhou, Huang, et~al\mbox{.}}{Huang et~al\mbox{.}}{2020}]%
        {tidb2020}
\bibfield{author}{\bibinfo{person}{Dongxu Huang}, \bibinfo{person}{Qi Liu},
  \bibinfo{person}{Qiu Cui}, \bibinfo{person}{Zhuhe Fang},
  \bibinfo{person}{Xiaoyu Ma}, \bibinfo{person}{Fei Xu}, \bibinfo{person}{Li
  Shen}, \bibinfo{person}{Liu Tang}, \bibinfo{person}{Yuxing Zhou},
  \bibinfo{person}{Menglong Huang}, {et~al\mbox{.}}}
  \bibinfo{year}{2020}\natexlab{}.
\newblock \showarticletitle{TiDB: a Raft-based HTAP database}.
\newblock \bibinfo{journal}{\emph{Proceedings of the VLDB Endowment}}
  \bibinfo{volume}{13}, \bibinfo{number}{12} (\bibinfo{year}{2020}),
  \bibinfo{pages}{3072--3084}.
\newblock


\bibitem[\protect\citeauthoryear{Huang, Cheng, Wang, Wang, He, Zhang, Li, Wang,
  Cao, and Li}{Huang et~al\mbox{.}}{2019}]%
        {Huang2019}
\bibfield{author}{\bibinfo{person}{Gui Huang}, \bibinfo{person}{Xuntao Cheng},
  \bibinfo{person}{Jianying Wang}, \bibinfo{person}{Yujie Wang},
  \bibinfo{person}{Dengcheng He}, \bibinfo{person}{Tieying Zhang},
  \bibinfo{person}{Feifei Li}, \bibinfo{person}{Sheng Wang},
  \bibinfo{person}{Wei Cao}, {and} \bibinfo{person}{Qiang Li}.}
  \bibinfo{year}{2019}\natexlab{}.
\newblock \showarticletitle{{X-engine: An optimized storage engine for
  large-scale e-commerce transaction processing}}. In
  \bibinfo{booktitle}{\emph{Proceedings of the ACM SIGMOD International
  Conference on Management of Data}}. \bibinfo{publisher}{Association for
  Computing Machinery}, \bibinfo{pages}{651--665}.
\newblock
\showISBNx{9781450356435}
\showISSN{07308078}


\bibitem[\protect\citeauthoryear{Jahangiri, Carey, and Freytag}{Jahangiri
  et~al\mbox{.}}{2022}]%
        {JahangiriCF22DJoin}
\bibfield{author}{\bibinfo{person}{Shiva Jahangiri},
  \bibinfo{person}{Michael~J. Carey}, {and} \bibinfo{person}{Johann{-}Christoph
  Freytag}.} \bibinfo{year}{2022}\natexlab{}.
\newblock \showarticletitle{Design Trade-offs for a Robust Dynamic Hybrid Hash
  Join}.
\newblock \bibinfo{journal}{\emph{Proc. {VLDB} Endow.}} \bibinfo{volume}{15},
  \bibinfo{number}{10} (\bibinfo{year}{2022}), \bibinfo{pages}{2257--2269}.
\newblock


\bibitem[\protect\citeauthoryear{Lahiri, Chavan, Colgan, Das, Ganesh, Gleeson,
  Hase, Holloway, Kamp, Lee, et~al\mbox{.}}{Lahiri et~al\mbox{.}}{2015}]%
        {oracle2015}
\bibfield{author}{\bibinfo{person}{Tirthankar Lahiri}, \bibinfo{person}{Shasank
  Chavan}, \bibinfo{person}{Maria Colgan}, \bibinfo{person}{Dinesh Das},
  \bibinfo{person}{Amit Ganesh}, \bibinfo{person}{Mike Gleeson},
  \bibinfo{person}{Sanket Hase}, \bibinfo{person}{Allison Holloway},
  \bibinfo{person}{Jesse Kamp}, \bibinfo{person}{Teck-Hua Lee},
  {et~al\mbox{.}}} \bibinfo{year}{2015}\natexlab{}.
\newblock \showarticletitle{Oracle database in-memory: A dual format in-memory
  database}. In \bibinfo{booktitle}{\emph{2015 IEEE 31st International
  Conference on Data Engineering}}. IEEE, \bibinfo{pages}{1253--1258}.
\newblock


\bibitem[\protect\citeauthoryear{Lamb, Fuller, Varadarajan, Tran, Vandiver,
  Doshi, and Bear}{Lamb et~al\mbox{.}}{2012}]%
        {lamb2012vertica}
\bibfield{author}{\bibinfo{person}{Andrew Lamb}, \bibinfo{person}{Matt Fuller},
  \bibinfo{person}{Ramakrishna Varadarajan}, \bibinfo{person}{Nga Tran},
  \bibinfo{person}{Ben Vandiver}, \bibinfo{person}{Lyric Doshi}, {and}
  \bibinfo{person}{Chuck Bear}.} \bibinfo{year}{2012}\natexlab{}.
\newblock \showarticletitle{The Vertica Analytic Database: C-Store 7 Years
  Later}.
\newblock \bibinfo{journal}{\emph{Proc. {VLDB} Endow.}} \bibinfo{volume}{5},
  \bibinfo{number}{12} (\bibinfo{year}{2012}), \bibinfo{pages}{1790--1801}.
\newblock


\bibitem[\protect\citeauthoryear{Larson, Birka, Hanson, Huang, Nowakiewicz, and
  Papadimos}{Larson et~al\mbox{.}}{2015}]%
        {larson2015real}
\bibfield{author}{\bibinfo{person}{Per-{\AA}ke Larson}, \bibinfo{person}{Adrian
  Birka}, \bibinfo{person}{Eric~N Hanson}, \bibinfo{person}{Weiyun Huang},
  \bibinfo{person}{Michal Nowakiewicz}, {and} \bibinfo{person}{Vassilis
  Papadimos}.} \bibinfo{year}{2015}\natexlab{}.
\newblock \showarticletitle{Real-time analytical processing with SQL server}.
\newblock \bibinfo{journal}{\emph{Proceedings of the VLDB Endowment}}
  \bibinfo{volume}{8}, \bibinfo{number}{12} (\bibinfo{year}{2015}),
  \bibinfo{pages}{1740--1751}.
\newblock


\bibitem[\protect\citeauthoryear{Larson, Clinciu, Hanson, Oks, Price,
  Rangarajan, Surna, and Zhou}{Larson et~al\mbox{.}}{2011}]%
        {larson2011sql}
\bibfield{author}{\bibinfo{person}{Per-{\AA}ke Larson}, \bibinfo{person}{Cipri
  Clinciu}, \bibinfo{person}{Eric~N Hanson}, \bibinfo{person}{Artem Oks},
  \bibinfo{person}{Susan~L Price}, \bibinfo{person}{Srikumar Rangarajan},
  \bibinfo{person}{Aleksandras Surna}, {and} \bibinfo{person}{Qingqing Zhou}.}
  \bibinfo{year}{2011}\natexlab{}.
\newblock \showarticletitle{SQL server column store indexes}. In
  \bibinfo{booktitle}{\emph{Proceedings of the 2011 ACM SIGMOD International
  Conference on Management of data}}. \bibinfo{pages}{1177--1184}.
\newblock


\bibitem[\protect\citeauthoryear{Lee, Moon, Kim, Kim, Cha, and Han}{Lee
  et~al\mbox{.}}{2017}]%
        {lee2017parallel}
\bibfield{author}{\bibinfo{person}{Juchang Lee}, \bibinfo{person}{SeungHyun
  Moon}, \bibinfo{person}{Kyu~Hwan Kim}, \bibinfo{person}{Deok~Hoe Kim},
  \bibinfo{person}{Sang~Kyun Cha}, {and} \bibinfo{person}{Wook-Shin Han}.}
  \bibinfo{year}{2017}\natexlab{}.
\newblock \showarticletitle{Parallel replication across formats in SAP HANA for
  scaling out mixed OLTP/OLAP workloads}.
\newblock \bibinfo{journal}{\emph{Proceedings of the VLDB Endowment}}
  \bibinfo{volume}{10}, \bibinfo{number}{12} (\bibinfo{year}{2017}),
  \bibinfo{pages}{1598--1609}.
\newblock


\bibitem[\protect\citeauthoryear{Leis, Boncz, Kemper, and Neumann}{Leis
  et~al\mbox{.}}{2014}]%
        {LeisBK014Morsel}
\bibfield{author}{\bibinfo{person}{Viktor Leis}, \bibinfo{person}{Peter~A.
  Boncz}, \bibinfo{person}{Alfons Kemper}, {and} \bibinfo{person}{Thomas
  Neumann}.} \bibinfo{year}{2014}\natexlab{}.
\newblock \showarticletitle{Morsel-driven parallelism: a NUMA-aware query
  evaluation framework for the many-core age}. In
  \bibinfo{booktitle}{\emph{International Conference on Management of Data,
  {SIGMOD} 2014, Snowbird, UT, USA, June 22-27, 2014}}.
  \bibinfo{publisher}{{ACM}}, \bibinfo{pages}{743--754}.
\newblock


\bibitem[\protect\citeauthoryear{Li, Dong, and Zhang}{Li et~al\mbox{.}}{2022}]%
        {CloudDB}
\bibfield{author}{\bibinfo{person}{Guoliang Li}, \bibinfo{person}{Haowen Dong},
  {and} \bibinfo{person}{Chao Zhang}.} \bibinfo{year}{2022}\natexlab{}.
\newblock \showarticletitle{Cloud Databases: New Techniques, Challenges, and
  Opportunities}.
\newblock \bibinfo{journal}{\emph{Proc. {VLDB} Endow.}} \bibinfo{volume}{15},
  \bibinfo{number}{12} (\bibinfo{year}{2022}), \bibinfo{pages}{3758--3761}.
\newblock


\bibitem[\protect\citeauthoryear{Li, Miao, Wu, Li, Wang, Cao, Qiao, Ruan,
  Liang, Yang, Dai, and Chen}{Li et~al\mbox{.}}{2023}]%
        {ROVEC}
\bibfield{author}{\bibinfo{person}{Meng Li}, \bibinfo{person}{Zheyu Miao},
  \bibinfo{person}{Di Wu}, \bibinfo{person}{Feifei Li}, \bibinfo{person}{Sheng
  Wang}, \bibinfo{person}{Wei Cao}, \bibinfo{person}{Zhi Qiao},
  \bibinfo{person}{Yubin Ruan}, \bibinfo{person}{Yukun Liang},
  \bibinfo{person}{Jimmy Yang}, \bibinfo{person}{Haipeng Dai}, {and}
  \bibinfo{person}{Guihai Chen}.} \bibinfo{year}{2023}\natexlab{}.
\newblock \showarticletitle{{ROVEC:} Runtime Optimization of Vectorized
  Expression Evaluation for Column Store}.
\newblock \bibinfo{journal}{\emph{{IEEE} Trans. Knowl. Data Eng.}}
  \bibinfo{volume}{35}, \bibinfo{number}{3} (\bibinfo{year}{2023}),
  \bibinfo{pages}{3045--3058}.
\newblock


\bibitem[\protect\citeauthoryear{Mishne, Dalton, Li, Sharma, and Lin}{Mishne
  et~al\mbox{.}}{2013}]%
        {Twitter}
\bibfield{author}{\bibinfo{person}{Gilad Mishne}, \bibinfo{person}{Jeff
  Dalton}, \bibinfo{person}{Zhenghua Li}, \bibinfo{person}{Aneesh Sharma},
  {and} \bibinfo{person}{Jimmy Lin}.} \bibinfo{year}{2013}\natexlab{}.
\newblock \showarticletitle{Fast data in the era of big data: Twitter's
  real-time related query suggestion architecture}. In
  \bibinfo{booktitle}{\emph{Proceedings of the 2013 ACM SIGMOD International
  Conference on Management of Data}}. \bibinfo{pages}{1147--1158}.
\newblock


\bibitem[\protect\citeauthoryear{Moerkotte and Neumann}{Moerkotte and
  Neumann}{2008}]%
        {MoerkotteN08}
\bibfield{author}{\bibinfo{person}{Guido Moerkotte} {and}
  \bibinfo{person}{Thomas Neumann}.} \bibinfo{year}{2008}\natexlab{}.
\newblock \showarticletitle{Dynamic programming strikes back}. In
  \bibinfo{booktitle}{\emph{Proceedings of the {ACM} {SIGMOD} International
  Conference on Management of Data, {SIGMOD} 2008, Vancouver, BC, Canada, June
  10-12, 2008}}. \bibinfo{publisher}{{ACM}}, \bibinfo{pages}{539--552}.
\newblock


\bibitem[\protect\citeauthoryear{MySQL.}{MySQL.}{2019}]%
        {mysql8}
\bibfield{author}{\bibinfo{person}{MySQL.}} \bibinfo{year}{2019}\natexlab{}.
\newblock \bibinfo{title}{{MySQL 8.0.18 (2019-10-14, General Availability).}}
\newblock
  \bibinfo{howpublished}{\url{https://dev.mysql.com/doc/relnotes/mysql/8.0/en/news-8-0-18.html}}.
\newblock


\bibitem[\protect\citeauthoryear{Ongaro and Ousterhout}{Ongaro and
  Ousterhout}{2014}]%
        {ongaro2014search}
\bibfield{author}{\bibinfo{person}{Diego Ongaro} {and} \bibinfo{person}{John
  Ousterhout}.} \bibinfo{year}{2014}\natexlab{}.
\newblock \showarticletitle{In search of an understandable consensus
  algorithm}. In \bibinfo{booktitle}{\emph{2014 USENIX Annual Technical
  Conference (Usenix ATC 14)}}. \bibinfo{pages}{305--319}.
\newblock


\bibitem[\protect\citeauthoryear{Oracle}{Oracle}{2018}]%
        {supplog}
\bibfield{author}{\bibinfo{person}{Oracle}.} \bibinfo{year}{2018}\natexlab{}.
\newblock \bibinfo{title}{Database-Level Supplemental Logging}.
\newblock
  \bibinfo{howpublished}{\url{https://docs.oracle.com/database/121/SUTIL/GUID-D2DDD67C-E1CC-45A6-A2A7-198E4C142FA3.htm}}.
\newblock


\bibitem[\protect\citeauthoryear{Pendse, Krishnaswamy, Kulkarni, Li, Lahiri,
  Raja, Zheng, Girkar, and Kulkarni}{Pendse et~al\mbox{.}}{2020}]%
        {pendse2020oracle}
\bibfield{author}{\bibinfo{person}{Sukhada Pendse}, \bibinfo{person}{Vasudha
  Krishnaswamy}, \bibinfo{person}{Kartik Kulkarni}, \bibinfo{person}{Yunrui
  Li}, \bibinfo{person}{Tirthankar Lahiri}, \bibinfo{person}{Vivekanandhan
  Raja}, \bibinfo{person}{Jing Zheng}, \bibinfo{person}{Mahesh Girkar}, {and}
  \bibinfo{person}{Akshay Kulkarni}.} \bibinfo{year}{2020}\natexlab{}.
\newblock \showarticletitle{Oracle database in-memory on active data guard:
  Real-time analytics on a standby database}. In \bibinfo{booktitle}{\emph{2020
  IEEE 36th International Conference on Data Engineering (ICDE)}}. IEEE,
  \bibinfo{pages}{1570--1578}.
\newblock


\bibitem[\protect\citeauthoryear{Prout, Wang, Victor, Sun, Li, Chen, Bergeron,
  Hanson, Walzer, Gomes, and Shamgunov}{Prout et~al\mbox{.}}{2022}]%
        {memsql2022}
\bibfield{author}{\bibinfo{person}{Adam Prout}, \bibinfo{person}{Szu{-}Po
  Wang}, \bibinfo{person}{Joseph Victor}, \bibinfo{person}{Zhou Sun},
  \bibinfo{person}{Yongzhu Li}, \bibinfo{person}{Jack Chen},
  \bibinfo{person}{Evan Bergeron}, \bibinfo{person}{Eric~N. Hanson},
  \bibinfo{person}{Robert Walzer}, \bibinfo{person}{Rodrigo Gomes}, {and}
  \bibinfo{person}{Nikita Shamgunov}.} \bibinfo{year}{2022}\natexlab{}.
\newblock \showarticletitle{Cloud-Native Transactions and Analytics in
  SingleStore}. In \bibinfo{booktitle}{\emph{{SIGMOD} '22: International
  Conference on Management of Data, Philadelphia, PA, USA, June 12 - 17,
  2022}}. \bibinfo{publisher}{{ACM}}, \bibinfo{pages}{2340--2352}.
\newblock


\bibitem[\protect\citeauthoryear{Raman, Attaluri, Barber, Chainani, Kalmuk,
  KulandaiSamy, Leenstra, Lightstone, Liu, Lohman, et~al\mbox{.}}{Raman
  et~al\mbox{.}}{2013}]%
        {raman2013db2}
\bibfield{author}{\bibinfo{person}{Vijayshankar Raman}, \bibinfo{person}{Gopi
  Attaluri}, \bibinfo{person}{Ronald Barber}, \bibinfo{person}{Naresh
  Chainani}, \bibinfo{person}{David Kalmuk}, \bibinfo{person}{Vincent
  KulandaiSamy}, \bibinfo{person}{Jens Leenstra}, \bibinfo{person}{Sam
  Lightstone}, \bibinfo{person}{Shaorong Liu}, \bibinfo{person}{Guy~M Lohman},
  {et~al\mbox{.}}} \bibinfo{year}{2013}\natexlab{}.
\newblock \showarticletitle{DB2 with BLU acceleration: So much more than just a
  column store}.
\newblock \bibinfo{journal}{\emph{Proceedings of the VLDB Endowment}}
  \bibinfo{volume}{6}, \bibinfo{number}{11} (\bibinfo{year}{2013}),
  \bibinfo{pages}{1080--1091}.
\newblock


\bibitem[\protect\citeauthoryear{Shen, Chen, Chen, and Zang}{Shen
  et~al\mbox{.}}{2021}]%
        {shen2021retrofitting}
\bibfield{author}{\bibinfo{person}{Sijie Shen}, \bibinfo{person}{Rong Chen},
  \bibinfo{person}{Haibo Chen}, {and} \bibinfo{person}{Binyu Zang}.}
  \bibinfo{year}{2021}\natexlab{}.
\newblock \showarticletitle{Retrofitting High Availability Mechanism to Tame
  Hybrid Transaction/Analytical Processing}. In \bibinfo{booktitle}{\emph{15th
  $\{$USENIX$\}$ Symposium on Operating Systems Design and Implementation
  ($\{$OSDI$\}$ 21)}}. \bibinfo{pages}{219--238}.
\newblock


\bibitem[\protect\citeauthoryear{Sikka, F{\"{a}}rber, Lehner, Cha, Peh, and
  Bornh{\"{o}}vd}{Sikka et~al\mbox{.}}{2012}]%
        {hana2012}
\bibfield{author}{\bibinfo{person}{Vishal Sikka}, \bibinfo{person}{Franz
  F{\"{a}}rber}, \bibinfo{person}{Wolfgang Lehner}, \bibinfo{person}{Sang~Kyun
  Cha}, \bibinfo{person}{Thomas Peh}, {and} \bibinfo{person}{Christof
  Bornh{\"{o}}vd}.} \bibinfo{year}{2012}\natexlab{}.
\newblock \showarticletitle{Efficient transaction processing in {SAP} {HANA}
  database: the end of a column store myth}. In
  \bibinfo{booktitle}{\emph{Proceedings of the {ACM} {SIGMOD} International
  Conference on Management of Data, {SIGMOD} 2012, Scottsdale, AZ, USA, May
  20-24, 2012}}. \bibinfo{publisher}{{ACM}}, \bibinfo{pages}{731--742}.
\newblock


\bibitem[\protect\citeauthoryear{SysBench}{SysBench}{2023}]%
        {sysbench}
\bibfield{author}{\bibinfo{person}{SysBench}.} \bibinfo{year}{2023}\natexlab{}.
\newblock \bibinfo{title}{{SysBench}}.
\newblock \bibinfo{howpublished}{\url{https://github.com/akopytov/sysbench}}.
\newblock


\bibitem[\protect\citeauthoryear{Vandiver, Prasad, Rana, Zik, Saeidi, Parimal,
  Pantela, and Dave}{Vandiver et~al\mbox{.}}{2018}]%
        {vandiver2018eon}
\bibfield{author}{\bibinfo{person}{Ben Vandiver}, \bibinfo{person}{Shreya
  Prasad}, \bibinfo{person}{Pratibha Rana}, \bibinfo{person}{Eden Zik},
  \bibinfo{person}{Amin Saeidi}, \bibinfo{person}{Pratyush Parimal},
  \bibinfo{person}{Styliani Pantela}, {and} \bibinfo{person}{Jaimin Dave}.}
  \bibinfo{year}{2018}\natexlab{}.
\newblock \showarticletitle{Eon Mode: Bringing the Vertica Columnar Database to
  the Cloud}. In \bibinfo{booktitle}{\emph{Proceedings of the 2018
  International Conference on Management of Data, {SIGMOD} Conference 2018,
  Houston, TX, USA, June 10-15, 2018}}. \bibinfo{publisher}{{ACM}},
  \bibinfo{pages}{797--809}.
\newblock


\bibitem[\protect\citeauthoryear{Vassiliadis}{Vassiliadis}{2009}]%
        {vassiliadis2009survey}
\bibfield{author}{\bibinfo{person}{Panos Vassiliadis}.}
  \bibinfo{year}{2009}\natexlab{}.
\newblock \showarticletitle{A survey of extract--transform--load technology}.
\newblock \bibinfo{journal}{\emph{International Journal of Data Warehousing and
  Mining (IJDWM)}} \bibinfo{volume}{5}, \bibinfo{number}{3}
  (\bibinfo{year}{2009}), \bibinfo{pages}{1--27}.
\newblock


\bibitem[\protect\citeauthoryear{Vera-Baquero, Colomo-Palacios, and
  Molloy}{Vera-Baquero et~al\mbox{.}}{2016}]%
        {BI}
\bibfield{author}{\bibinfo{person}{Alejandro Vera-Baquero},
  \bibinfo{person}{Ricardo Colomo-Palacios}, {and} \bibinfo{person}{Owen
  Molloy}.} \bibinfo{year}{2016}\natexlab{}.
\newblock \showarticletitle{Real-time business activity monitoring and analysis
  of process performance on big-data domains}.
\newblock \bibinfo{journal}{\emph{Telematics and Informatics}}
  \bibinfo{volume}{33}, \bibinfo{number}{3} (\bibinfo{year}{2016}),
  \bibinfo{pages}{793--807}.
\newblock


\bibitem[\protect\citeauthoryear{Verbitski, Gupta, Saha, Brahmadesam, Gupta,
  Mittal, Krishnamurthy, Maurice, Kharatishvili, and Bao}{Verbitski
  et~al\mbox{.}}{2017}]%
        {Aurora}
\bibfield{author}{\bibinfo{person}{Alexandre Verbitski},
  \bibinfo{person}{Anurag Gupta}, \bibinfo{person}{Debanjan Saha},
  \bibinfo{person}{Murali Brahmadesam}, \bibinfo{person}{Kamal Gupta},
  \bibinfo{person}{Raman Mittal}, \bibinfo{person}{Sailesh Krishnamurthy},
  \bibinfo{person}{Sandor Maurice}, \bibinfo{person}{Tengiz Kharatishvili},
  {and} \bibinfo{person}{Xiaofeng Bao}.} \bibinfo{year}{2017}\natexlab{}.
\newblock \showarticletitle{Amazon Aurora: Design Considerations for High
  Throughput Cloud-Native Relational Databases}. In
  \bibinfo{booktitle}{\emph{Proceedings of the 2017 {ACM} International
  Conference on Management of Data, {SIGMOD} Conference 2017, Chicago, IL, USA,
  May 14-19, 2017}}. \bibinfo{publisher}{{ACM}}, \bibinfo{pages}{1041--1052}.
\newblock


\bibitem[\protect\citeauthoryear{Verbitski, Gupta, Saha, Corey, Gupta,
  Brahmadesam, Mittal, Krishnamurthy, Maurice, Kharatishvilli,
  et~al\mbox{.}}{Verbitski et~al\mbox{.}}{2018}]%
        {verbitski2018amazon}
\bibfield{author}{\bibinfo{person}{Alexandre Verbitski},
  \bibinfo{person}{Anurag Gupta}, \bibinfo{person}{Debanjan Saha},
  \bibinfo{person}{James Corey}, \bibinfo{person}{Kamal Gupta},
  \bibinfo{person}{Murali Brahmadesam}, \bibinfo{person}{Raman Mittal},
  \bibinfo{person}{Sailesh Krishnamurthy}, \bibinfo{person}{Sandor Maurice},
  \bibinfo{person}{Tengiz Kharatishvilli}, {et~al\mbox{.}}}
  \bibinfo{year}{2018}\natexlab{}.
\newblock \showarticletitle{Amazon aurora: On avoiding distributed consensus
  for i/os, commits, and membership changes}. In
  \bibinfo{booktitle}{\emph{Proceedings of the 2018 International Conference on
  Management of Data}}. \bibinfo{pages}{789--796}.
\newblock


\bibitem[\protect\citeauthoryear{Yang, Rae, Xu, Shute, Yuan, Lau, Zeng, Zhao,
  Ma, Chen, et~al\mbox{.}}{Yang et~al\mbox{.}}{2020}]%
        {f12020}
\bibfield{author}{\bibinfo{person}{Jiacheng Yang}, \bibinfo{person}{Ian Rae},
  \bibinfo{person}{Jun Xu}, \bibinfo{person}{Jeff Shute}, \bibinfo{person}{Zhan
  Yuan}, \bibinfo{person}{Kelvin Lau}, \bibinfo{person}{Qiang Zeng},
  \bibinfo{person}{Xi Zhao}, \bibinfo{person}{Jun Ma}, \bibinfo{person}{Ziyang
  Chen}, {et~al\mbox{.}}} \bibinfo{year}{2020}\natexlab{}.
\newblock \showarticletitle{F1 Lightning: HTAP as a Service}.
\newblock \bibinfo{journal}{\emph{Proceedings of the VLDB Endowment}}
  \bibinfo{volume}{13}, \bibinfo{number}{12} (\bibinfo{year}{2020}),
  \bibinfo{pages}{3313--3325}.
\newblock


\bibitem[\protect\citeauthoryear{Zhan, Su, Wei, Peng, Lin, Wang, Chen, Li, Pan,
  Zheng, and Chai}{Zhan et~al\mbox{.}}{2019}]%
        {DBLP:journals/pvldb/ZhanSWPLWCLPZC19}
\bibfield{author}{\bibinfo{person}{Chaoqun Zhan}, \bibinfo{person}{Maomeng Su},
  \bibinfo{person}{Chuangxian Wei}, \bibinfo{person}{Xiaoqiang Peng},
  \bibinfo{person}{Liang Lin}, \bibinfo{person}{Sheng Wang},
  \bibinfo{person}{Zhe Chen}, \bibinfo{person}{Feifei Li}, \bibinfo{person}{Yue
  Pan}, \bibinfo{person}{Fang Zheng}, {and} \bibinfo{person}{Chengliang Chai}.}
  \bibinfo{year}{2019}\natexlab{}.
\newblock \showarticletitle{AnalyticDB: Real-time {OLAP} Database System at
  Alibaba Cloud}.
\newblock \bibinfo{journal}{\emph{Proc. {VLDB} Endow.}} \bibinfo{volume}{12},
  \bibinfo{number}{12} (\bibinfo{year}{2019}), \bibinfo{pages}{2059--2070}.
\newblock


\bibitem[\protect\citeauthoryear{Zhou, Li, Zhao, Chen, Li, Yang, Cui, Yu, Chen,
  Ding, and Qi}{Zhou et~al\mbox{.}}{2017}]%
        {KunPeng}
\bibfield{author}{\bibinfo{person}{Jun Zhou}, \bibinfo{person}{Xiaolong Li},
  \bibinfo{person}{Peilin Zhao}, \bibinfo{person}{Chaochao Chen},
  \bibinfo{person}{Longfei Li}, \bibinfo{person}{Xinxing Yang},
  \bibinfo{person}{Qing Cui}, \bibinfo{person}{Jin Yu}, \bibinfo{person}{Xu
  Chen}, \bibinfo{person}{Yi Ding}, {and} \bibinfo{person}{Yuan~Alan Qi}.}
  \bibinfo{year}{2017}\natexlab{}.
\newblock \showarticletitle{KunPeng: Parameter Server Based Distributed
  Learning Systems and Its Applications in Alibaba and Ant Financial}. In
  \bibinfo{booktitle}{\emph{Proceedings of the 23rd ACM SIGKDD International
  Conference on Knowledge Discovery and Data Mining}}
  \emph{(\bibinfo{series}{KDD '17})}. \bibinfo{publisher}{Association for
  Computing Machinery}, \bibinfo{address}{New York, NY, USA},
  \bibinfo{pages}{1693–1702}.
\newblock
\showISBNx{9781450348874}


\end{thebibliography}

\end{document}